# A resilience glossary shaped by context: Reviewing resilience-related terms for critical infrastructures

Andrea Mentges [a,*], Lukas Halekotte [a,+,*], Moritz Schneider [a], Tobias Demmer [a], and Daniel Lichte [a]

[a] *German Aerospace Center (DLR),*
*Institute for the Protection of Terrestrial Infrastructures,*
*Rathausallee 12, 53757, Sankt Augustin, Germany*

[+] Corresponding author.
*E-mail address:* lukas.halekotte@dlr.de

* Co-first authors.

**Abstract**

We present a comprehensive resilience glossary, comprising a set of 91 definitions of resilience-related terms used in the context of critical infrastructures. The definition and use of many of these terms, as well as the term resilience itself, shows an enormous variability in the literature. Therefore, we draw from the diverse pool of published definitions, integrate multiple contrasting views, compare the individual terms, and provide references to adjoining or contesting views, to create a clear resilience terminology. This terminology outlines a specific understanding of resilience which supports the effective assessment and management of the resilience of critical infrastructures. The two central elements of this understanding are that (1) resilience is the ability of a system to deal with the impacts of unspecific and possibly unforeseen disruptive events, and that (2) this ability comprises three pillar capacities whose quality can be extracted from performance curves.

*Keywords:* Resilience definition; Critical infrastructures; Terminology; Resilience capacities; Performance curve

# 1 Introduction

## 1.1 Background

Sometimes, failure is inevitable. For more than 50 years, this notion pushed scientists from various fields to study the ability of systems and individuals to deal with sudden negative impacts (Comfort et al., 2010). Today, understanding this ability – famously known under the name resilience - is more relevant than ever. Resilience can help critical infrastructures to master challenges posed by various global changes, such as globalization, digitalization, and the growing number of extreme weather events (Bie et al., 2017; Caralli et al., 2010; Comfort et al., 2010; Thier & Pot d'Or, 2020); and thus, protect the essential services and assets our societies rely on.

## 1.2 Motivation

The definition of resilience shows an enormous variability. The long history of the concept (Alexander, 2013), paired with early and persistent disagreements on its best use (see, e.g., the differing perspectives of Holling (1973) and Pimm (1984)), and its adaptation and continuous development in various research fields (e.g., ecology (S. Carpenter et al., 2001; Holling, 1973; Pimm, 1984), disaster management (Adger et al., 2005; Bruneau et al., 2003; Cutter et al., 2008) or critical infrastructure protection (Guo et al., 2021; Poulin & Kane, 2021)) have led to its diversification. As a result, the exact definition of resilience varies across domains (Alexander, 2013; Xue et al., 2018) and within domains (Asadzadeh et al., 2017; Rus et al., 2018), in dependence on research focus (e.g., seismic resilience of communities (Bruneau et al., 2003), hurricane resilience of electric power systems (Ouyang & Dueñas-Osorio, 2014) or urban climate change resilience (A. Brown et al., 2012)), and from author to author (see for example the extensive collection of definitions in Wied et al. (2020), Biringer et al. (2013), or Carlson et al. (2012)). Therefore, there is no one-size-fits-all definition of resilience (Jore, 2020) and concluding consensus about its definition has not been reached (Mottahedi et al., 2021). A clear definition is crucial, however, since the way we define resilience will shape the way we measure it (Biringer et al., 2013). The aim of this document is to compile, define, and distinguish terms that are used to describe the resilience of socio-technical systems, and in particular of critical infrastructures. The goal is to collect synonyms and carve out subtle differences among similar terms, as well as to provide references and resources to gain further insights about each term.

## 1.3 What has been done before

Other resilience glossaries have been published before. However, they are mostly not targeted at the context of critical infrastructures, but have a different focus, e.g., the Society for Risk Analysis glossary (Aven et al., 2018), the industry-centered glossary by DRI International (DRI, 2021), the informatics-centered glossary by Andersson and colleagues (Andersson et al., 2020), or the ecological resilience glossary by the Resilience Alliance (Resilience Alliance, 2021). A recent study reviewed 17 definitions of terms which are described as contributing to infrastructure resilience (Mottahedi et al., 2021). The unique aspect of this glossary is the emphasis on carving out and discussing the differences and similarities between various terms which are used in the context of infrastructure resilience.

## 1.4 Aim of the study

We define and distinguish more than 90 resilience-related terms in the context of critical infrastructures. Just like the term resilience, the exact meaning of many of these terms is disputable and changes depending on the research field. In this regard, we emphasize that we do not claim to have found the "right" definition nor that others' definitions are flawed. Instead, we built on previous work to find an integrated terminology, that comprises many resilience-related terms without duplications and as much distinctness between them as possible. The aim of this work is thus not to add yet another definition to the comprehensive, valuable work that has already been done to define resilience and its related concepts. Rather, we aim to bring the existing definitions together, integrate them, condense them, and obtain a consistent set of terms which are as clearly distinct from each other as possible. Therefore, in the following, we i) briefly summarize the status of resilience definition, ii) show that there are still contradictions, blurriness, and overlap between terms, iii) present a glossary of resilience-related terms based on existing literature, describing the meaning of each term and the differences between them, and iv) discuss the implications of our view on resilience for its quantification and management.

## 2 Current status of the resilience definition

The concept of resilience has been established in the context of critical infrastructures mainly in the past two decades and the number of related studies is still growing significantly (Mottahedi et al., 2021).

### 2.1 Variability in terminology: resilience

The definition of resilience depends on the author. Several previous studies have compiled extensive lists of resilience definitions from numerous sources, concluding that the variation in definitions is high (e.g., Mottahedi et al. (2021) enumerate 28 definitions; Wied et al. (2020) compare 251 definitions; see also, Biringer et al. (2013), or Carlson et al. (2012)). Due to the great number of definitions and the high variability among these definitions, the concept of resilience has been heavily criticized. It has been claimed, for instance, that the term has a "poor scientific status" in regard to its application in terrorism research (Jore, 2020), that it is "potentially disappointing to read too much into the term as a model and as a paradigm" (Alexander, 2013), that the term is a buzzword (Linkov et al., 2014), and might be used more broadly than it should (Rose, 2007). In fact, it is not always clear what the term resilience refers to. Many authors describe resilience as an ability or capability of a system (B. Cai et al., 2018; Engler et al., 2020; Francis & Bekera, 2014; Ouyang & Wang, 2015; Rose, 2007; Sterbenz et al., 2010). However, resilience is sometimes also understood as the process or outcome following the disruption of a system (Béné et al., 2012); an understanding which allows a much easier assessment. Some authors state that both the "power or ability" (i.e., capacities) and the "action or act" (i.e., processes) are included in resilience (Kanno et al., 2019). Other well-known definitions describe resilience not conceptually as a capacity or process, but as a measure, e.g., a speed, rate, or degree of something; for example, the definition of engineering resilience according to Pimm (1984), which emphasizes the speed of recovery. However, several reviews of resilience definitions from the engineering and infrastructure context (Biringer et al., 2013; Mottahedi et al., 2021; Wied et al., 2020) show that most studies view resilience as a capacity. Here, we follow this view, seeing the resilience processes as a result of these capacities in the case of a disruption. Therefore, we do not consider the processes a part of the resilience term, but restrict our view of resilience to the capacity.

### 2.2 Variability in terminology: capacities

Resilience is an umbrella term, which comprises many different system capacities (sometimes termed components (Carlson et al., 2012; Rehak et al., 2019) or dimensions (Bruneau et al., 2003)). Most sources (see Wied et al. (2020)) emphasize the ability to recover (Teodorescu, 2015), some the ability to absorb (Vugrin et al., 2011), or learn (Folke, 2016; NIAC, 2010; Norris et al., 2008). Like for the term resilience itself, the definitions of its capacities are highly variable. For example, Carlson and colleagues' (2012) definition of mitigation includes resistance and absorptivity while Ouyang and colleagues' (2012) definition of resistance describes prevention and reduction. In addition, the grouping and number of resilience capacities varies across studies. For example, Béné et al. (2012) distinguish absorptive, adaptive, and transformative capacity; Rehak et al. (2019) distinguish robustness, recoverability, and adaptability; the MCEER's resilience framework "4R" comprises robustness, redundancy, resourcefulness, and rapidity (Bruneau et al., 2003); and the Argonne National Laboratory's framework comprises preparedness, mitigation, response, and recovery (Carlson et al., 2012).

Furthermore, the meaning of individual capacities also varies from author to author. An illustrative example is the adaptive capacity whose definition varies depending on, among others, the type of disruption which is considered, the time at which this capacity becomes relevant (i.e., during or after the disruption) as well as its effect (i.e., sustain performance or constrain performance loss (Béné et al., 2012; Francis & Bekera, 2014), restore performance (Nan et al., 2014), enhance performance (Häring et al., 2017), improve resilience (NIAC, 2010; Rehak et al., 2019)). For instance, Béné et al. (2012) assume that the adaptive capacity of a system becomes important once the intensity of a disruption (in this case, a press disturbance) exceeds the system's absorptive capacity – adaptive capacity enables the maintenance of system functioning in the light of greater persistent disruptions. By contrast, other authors describe the adaptive capacity as the ability to draw lessons from past disruptions in order to improve the remaining resilience capacities, i.e., adaptive capacity plays out mainly after the restoration (NIAC, 2010; Rehak et al., 2019). In this sense, if adaptability is high, resilience is a self-reinforcing ability.

## 2.3 Variability in terminology: phases

Similarly, there is considerable variation in terminology concerning the phases of the resilience cycle. The resilience cycle represents the behavior of a system in the form of an idealized performance curve, which typically starts shortly before a single disruptive event and ends after the system has recovered from the associated impacts. The curve is conceptually divided into successive phases, which are named differently depending on the author (Figure 1). Most authors name the phases in accordance with the resilience process that takes place during that time (Braun et al., 2020; Jovanović et al., 2020; Klimek et al., 2019). Some refer to the resilience capacity that is thought to be most important during the corresponding time period (Rehak et al., 2018). Others distinguish the phases based on the state of the system or the progression of the disruption (Hossain et al., 2019; Mottahedi et al., 2021). However, despite all these differences in the naming, there is a consensus on the general order and duration of the phases (Figure 1).

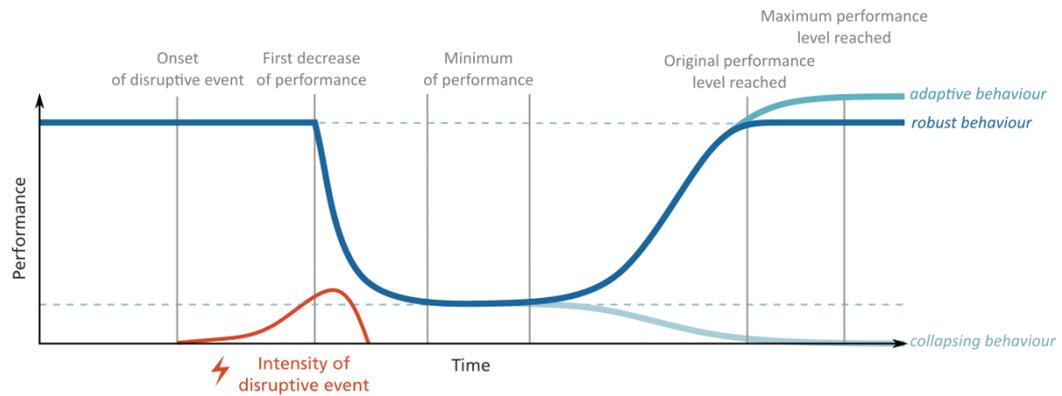

**Figure 1: Resilience phases across studies.** Depending on the author, the phases of the resilience cycle are called differently, emphasizing either the current state of the system, a resilience capacity that is important during the respective phase, or a resilience process that typically takes place at the time. However, the listed examples illustrate that start and end time points of the phases are largely consistent across studies.

## 2.4 What's *not* resilience?

To understand the concept of resilience, it can be helpful to review its boundaries, i.e., what is *not* resilience. There are many different perspectives on where the resilience concept ends. Holling (1973), the originator of the ecological resilience concept, emphasizes the difference between resilience and stability (similar to Norris et al. (2008)), where stability is defined as the ability of a system to return to its original equilibrium state after a temporary disruption, and resilience "is the ability of systems to absorb change and disruption and still maintain the same relationships between […] state variables" (Holling, 1973). Thus, Holling highlights the aspect of flexibility and transformability that is included in the resilience concept. While stability and resilience can be understood as complementary abilities of a system, other authors contrast resilience with its supposable opposite to illustrate the concept. Woods (2015, 2016) sees resilience as the opposite of brittleness, with brittleness describing the sudden collapse of a system at or beyond its operating range, and resilience describing the ability to persist in the face of surprising disruptions (i.e., graceful extensibility). In social sciences, fragility is sometimes seen as the opposite of resilience (Manyena & Gordon, 2015; OECD, 2002). Also, vulnerability is sometimes treated as the opposite of resilience (Adger et al., 2005) or can also be seen as the opposite of the resilience capacity robustness (Mottahedi et al., 2021), since robustness is often measured as the inverse of the amount of performance loss (Shinozuka et al., 2004), while vulnerability is often defined as the amount of performance loss (Aven et al., 2018; Fischer et al., 2018; Häring et al., 2017). In our view, there is no veritable opposite to the resilience concept. Instead, high resilience entails the absence or minimization of, among others, vulnerability, fragility, and inflexibility.

An important difference lies between resilience management and risk management. The differentiation and relationship between the two is vividly discussed (Aven, 2019; Fekete et al., 2020; Linkov et al., 2018; Linkov & Palma-Oliveira, 2017), with opposing views ranging from refusing that resilience management adds anything new to existing risk practices to the believe that it represents a compulsory addition to the former. We see them as two complementary approaches for dealing with disruptions (see also Discussion section 4.1.3), which can benefit from each other in the joint goal of protecting infrastructures, as others have postulated before (Johansson et al., 2013; Park et al., 2013). Risk management focuses on identifying and avoiding or reducing the impact of foreseeable specific threats (Ganin et al., 2016; Park et al., 2013). While risk management might accept/acknowledge the continuance of some residual risk, resilience management is explicitly motivated by the notion that neither are all risks predictable nor is all risk avoidable (Linkov et al., 2018; Park et al., 2013). Thus, resilience management puts less emphasis on the reduction of individual risks, but focuses on a holistic increase of the ability to deal with disruptions as they emerge (Anholt & Boersma, 2018), emphasizing the time after the initial disruptive event, i.e., the recovery, learning, and adaption processes (Gasser et al., 2019).

## 3 Glossary

The following glossary presents a consistent terminology, that integrates various aspects of resilience and provides a distinction between them. We decided to order the terms by topic, rather than alphabetically, in order to create a document which can be read from front to back and thus can also serve as a primer for scholars who are rather new to the topic. The glossary begins with a definition of basic terms, such as system, disruption and risk (section 3.1); followed by terms that describe concepts which are important for system maintenance (section 3.2). In the following, we define both capacities that are important for traditional risk management (section 3.3) and capacities that are emphasized by resilience management (sections 3.4-3.6). The capacities are grouped according to the time when they become relevant, i.e., before the disruption (section 3.3), at the beginning of a disruption (section 3.4), immediately after a disruption (section 3.5), and in-between disruptions (section 3.6). Finally, we compile terms related to resilience management (section 3.7) and analysis (section 3.8). Throughout the glossary, indentions point out the hierarchy of terms: left aligned terms are broader umbrella terms, which include the indented sub-terms grouped below them. Synonyms are included in the descriptions of the terms.

### 3.1 Basic terms
The following terms are the basis to understand the resilience-related concepts.

| Term | Description |
|---|---|
| **System** | A set of components that act together as a whole to accomplish a specific function or set of functions (IEEE Std 610.12-1990; Skyttner, 2006). System components can be for example machines and cables in technical systems, or individuals with specific responsibilities and tasks in socio-technical systems. We acknowledge that, for the general definition of a system, the need to perform a function might not be an absolutely necessary condition (Backlund, 2000), but it is one which is very essential in the context of resilience: Only if a desired or acceptable behavior or state can be defined, i.e., one in which a certain function is fulfilled, there is something to be maintained or restored and thus the system can show resilient behavior. Accordingly, our definition corresponds to a system which meets the fundamental requirements to be resilient. |
| **System of systems** | A set of systems, which act together in a specific context to achieve a common goal, which none of the constituent systems could reach on its own (ISO, 2018; Nielsen et al., 2015). For example, a system of systems in the medical context is the health care system, where the emergency treatment of patients relies on the interaction of several distinct systems, i.e., hospital management, power, traffic (road system) and communication systems. |
| **Performance** | The status of a system with regard to its designated function(s), i.e., how well a system's current behavior fulfills its purpose. The term "performance" is often used as a synonym for its quantitative representation, i.e., the measure of performance ("the vertical axis of a resilience curve" (Poulin & Kane, 2021), see Figure 1). In accordance with the different functions a system can serve, performance and its measures can refer to different things. For instance, in the context of infrastructures, Poulin and Kane (2021) distinguish three different types of performance measures: Availability (e.g., number of functional pipes), productivity (e.g., water demand satisfied) and quality measures (e.g., water quality index).<br>*Synonym:* Functionality (Cimellaro et al., 2010)<br>*Measured by*: Performance variable (Wied et al., 2020), performance measure (Poulin & Kane, 2021), performance indicator, key performance indicator (Torres et al., 2020) |
| **Hazard, threat** | A potential state of the system or a situation which could cause damage or decrease functionality (Aven et al., 2018). Although not limited to, "hazard" often refers to natural and unintentional sources of danger or risk, such as a fire hazard, environmental hazard or accident hazard (safety-related); whereas "threat" more often refers to risks imposed by intentionally acting agents (security-related), i.e., a threat depends on intent and capability of the acting agents (Bieder & Pettersen Gould, 2020; Smith & Brooks, 2013). |
| **Exposure** | The degree to which a system or an asset is subject to a risk source (Adger, 2006; Aven et al., 2018). It is used in absolute, i.e., a system is either exposed or not exposed to a certain hazard or threat, as well as relative terms, i.e., a system can have a high or low exposure to a certain hazard or threat. For instance, in the context of disaster risk reduction, exposure can be expressed by the number of people, the |
| | situation of infrastructures, or the amount of property and other human assets which are present in a hazard-prone area and thus portray the potential for loss (Kreibich et al., 2022; UNISDR, 2009). |
| **Stressor** | All internal and external stimuli that cause stress and thereby lead the affected system to initiate an active reaction. Stressors can trigger disruptive events, and thus a loss of performance. The term "stressor" is well-defined and frequently used in psychology and ecology. For example, in humans, stressors such as prolonged noise can limit attention, which may lead to false decisions (Endsley, 1995). In contrast, in the context of infrastructure resilience, the usage of the term varies. Some authors see stressors as a synonym of disruptive events (Aven et al., 2018; Parsons et al., 2016), whereas others define stressor as a stimulus that may trigger events, for example climate change or increasing globalization (O'Brien et al., 2004). The latter definition corresponds to the original definition of the term in psychology and is therefore reflected in this glossary.<br>*Synonym*: Stress factor<br>*Examples*: Wind, climate change, scarcity of critical resources, material fatigue |
| **Incident** | An event or the occurrence of something with the potential for negative consequences (i.e., a potentially disruptive event). In the case of an incidence, a certain threat or hazard is realized which results in a specific exposure of the system. In contrast to a disruptive event, an incident does not necessarily lead to any negative consequences (e.g., the exposure could be zero or the system could fully absorb the impact). The term is often further specified in order to highlight its type (e.g., poisoning incident, terrorist bombing incident, mass casualty incident) or severity (e.g., major, serious, catastrophic incident) (Aylwin et al., 2006; National Academy of Sciences, 2012). Based on the cause of an incident, we can distinguish three general types of events (Biringer et al., 2013; Mitroff & Alpaslan, 2003): Natural hazardous events, accidents and attacks.<br>*Synonym:* Event (Wied et al., 2020) |
| **Natural hazardous event** | An incidence that has a natural (non-human) source, i.e., it is neither directly caused by humans nor indirectly by a human-made system (safety-related). This involves climate- and weather-related (e.g., floods, storms, droughts), geophysical (e.g., earthquakes, volcanic eruptions) and biological (e.g., disease outbreaks, insect infestations) incidents (IFRC, 2020).<br>*Synonym:* Natural accident (Mitroff & Alpaslan, 2003) |
| **Accident** | An incidence that is directly or indirectly caused by humans but happens without planning, for instance, as a consequence of an unintentional or erroneous human action, a technical malfunction or a system overload (safety-related). An accident usually, though not necessarily, has negative consequences such as damage, injury or pollution (N. G. Leveson, 2016).<br>*Synonym:* Normal accident (Mitroff & Alpaslan, 2003; Perrow, 1999) |
| **Attack** | An incidence that is caused by an intentional act (security-related). In contrast to a natural hazardous event or an accident, an attack is planned and executed by a responsive antagonist (e.g., a terrorist, criminal or cyber-attacker) who proceeds in accordance with its own |

| | | |
|---|---|---|
| | | capabilities and with regard to a specific purpose/intent (G. Brown et al., 2006; Haimes, 2006; Jenelius et al., 2010).<br>*Synonym:* Abnormal accident (Mitroff & Alpaslan, 2003), malevolent event (Biringer et al., 2013) |
| **Disruptive event** | | An event that leads to negative consequences, i.e., a disruptive event is the cause of a loss of performance (see Figure 1). In contrast to the more general term incident this implies that there is an actual (i.e., non-zero) exposure to the event and that the system is not able to fully absorb its impact. Comprises all sources of performance decrease, including, e.g., natural hazards, technical failures, human errors, extreme loads and organizational issues as well as intentional malicious attacks (Gasser et al., 2019; Häring et al., 2016; Z. Xu et al., 2020).<br>*Synonyms*: Disturbance (Pimm, 1984), adverse event (National Academy of Sciences, 2012), undesired event (Biringer et al., 2013) |
| | **Sudden-onset disruptive event** | A disruptive event whose intensity builds up instantaneously (Zobel & Khansa, 2012), e.g., an earthquake or a tidal flood. In the context of resilience, many authors focus on such sudden-onset disruptive events (Gasser et al., 2019; Klise et al., 2017). It refers to one of the two extremes regarding the onset speed of disruptive events (Hewitt & Burton, 1971), the other extreme being included in the class of slow-onset disruptive events. |
| | **Slow-onset disruptive event** | A disruptive event whose intensity increases over a longer period of time (Zobel & Khansa, 2012), e.g., deterioration of components or increasingly harsher conditions due to climatic changes. It should be noted that a slow-onset disruptive event can still lead to a sudden collapse of performance, e.g., when approaching a bifurcation or tipping point (Dai et al., 2012; Gao et al., 2016). We propose that whether the onset of a disruptive event is considered slow or fast should depend on the time-scale of the considered system, i.e., if the intrinsic dynamics of the system are faster than the emergence of the event, the event should be referred to as slow-onset.<br>*Synonym*: Gradual onset (Hewitt & Burton, 1971), creeping disruptive event (Häring et al., 2017), slow-onset changes or processes (van der Geest & van den Berg, 2021) |
| | **Pulse disturbance** | A (sudden-onset) disruptive event of extremely short duration, such as a chemical spill (Donohue et al., 2016). It refers to one of the two extremes regarding the duration spectrum of disruptive events, the other extreme being the press disturbance.<br>*Synonyms:* Shock (Ives & Carpenter, 2007), acute disturbance (Donohue et al., 2016) |
| | **Press disturbance** | A disruptive event that shows lasting impact (Donohue et al., 2016; Ives & Carpenter, 2007), such as a steady outflow of toxic chemicals. The term is mainly used in the ecological literature.<br>*Synonyms*: Adverse change (Wied et al., 2020), long-term change, chronic disturbance (Donohue et al., 2016) |
| | **High Impact Low Probability (HILP) event** | A disruptive event that is extreme, i.e., it is rare and causes severe consequences (Aven, 2015b; Mahzarnia et al., 2020; Masys, 2012; Panteli et al., 2017). |

| | |
|---|---|
| | *Synonym:* Extreme event (Mahzarnia et al., 2020; Panteli & Mancarella, 2015b; Umunnakwe et al., 2021), high-impact rare (HR) event (Gholami et al., 2018) |
| **Disruption** | A disruption is caused by a disruptive event and lasts as long as the performance of the system is decreased as a consequence of the event (i.e., a disruption is characterized by the performance drop in the resilience curve, see Figure 1). The disruption ends when the system has recovered, i.e., it has returned to a normal operating state (see also restorative capacity in section 3.5). Thus, the disruption might last longer than the disruptive event (or be less persistent, e.g., if a system is able adapt to a long-lasting disruptive event). An important implication of this definition is that the manifestation of a disruption depends on the causing event (disruptive event) and on capacities of the affected system (i.e., its resilience and vulnerability). The proposed distinction between disruption and disruptive event is in line with the use of the terms in (Rehak et al., 2016; Rehak et al., 2019; Roege et al., 2014), to name a few. It should, however, be noted that the two terms are often used synonymously and sometimes even diametrical to our definition (see, e.g., the use of disruptive event in Fox-Lent et al. (2015)). |
| **Disaster** | A disruption of the functioning of a community which tests or exceeds its capacity to cope with the situation using its own resources. It is characterized by severe and often widespread impacts on human well-being, e.g., death, disease, injury, and loss of property, essential services or natural resources (IFRC, 2022; UNISDR, 2009; United Nations General Assembly, 2016). Although natural events (e.g., storms, floods, earthquakes) are the most common cause, disasters can also be human-induced (e.g., transport or industrial accidents) (CRED, 2022; IFRC, 2020). It is important to note that a disaster does not correspond to its triggering (disruptive) event but is a function of the event and properties (vulnerability, exposure, capabilities) of the affected community and the involved actors (a disaster is a disruption not a disruptive event). |
| **Crisis** | Similar to disaster (see above), it describes a disruption during which the core services provided by a system are endangered and/or its performance is already significantly affected. For instance, the temporary or partial breakdown of a critical infrastructure which exceeds known patterns can be denoted as a crisis. The terms disaster and crisis are often used interchangeable. However, the term crisis generally refers to a less severe situation, e.g., one without fatalities or the permanent loss of critical services (Boin & McConnell, 2007). Nevertheless, if not properly dealt with, a crisis can turn into a disaster (UNISDR, 2009).<br>Defining crisis as a disruption is in line with the view of "crisis as a process" as outlined in Williams et al. (2017). This view implies that the response to the causing event shapes the emerging crisis. The contrary view would be to consider crisis as the causing event (i.e., the incident or disruptive event) which would align with the view of "crisis as an event" (Williams et al., 2017). |
| **Risk** | The concept of risk describes "uncertain exposure to perceived harm" (Smith & Brooks, 2013) or "the possibility of loss" (O'Neill, 2017). It |
| | describes both the probability of a harmful effect occurring and the expected magnitude of the undesirable consequence (UNDRR, 2019). Risk can be probabilistically formalized in different ways, for instance, as the product of probability and consequence (Smith & Brooks, 2013) or, in the context of security, the product of threat, vulnerability, and consequence (McGill et al., 2007), where vulnerability describes the likelihood of a negative impact of the disruptive event. Independent of the applied metric, risk can generally be described as the result of its (three) essential determinants: The risk source (i.e., the hazard or threat), and the exposure and vulnerability of the system to this risk source (Francis & Bekera, 2014; Lavell et al., 2012). |
| **Uncertainty** | Uncertainty can be described as "any departure from the unachievable ideal of complete determinism" (W. E. Walker et al., 2003). As such, it is more than the absence of knowledge (W. E. Walker et al., 2013). The perceived degree of uncertainty is partially subjective, as it depends on the satisfaction with the existing knowledge (W. E. Walker et al., 2013). In a socio-technical system, different sources of uncertainty can exist, for instance, in decision making, uncertainty can refer to the uncertain future state of the world as well as to the uncertainty regarding the behavior of different actors (Quade, 1989). |
| **Epistemic uncertainty** | A type of uncertainty which becomes deterministic over time as more data and/or knowledge are gathered. Thus, there is the possibility to reduce epistemic uncertainty by gathering more data or by refining models (Kiureghian & Ditlevsen, 2009; W. E. Walker et al., 2003). |
| **Aleatory uncertainty** | A type of uncertainty where the modeler does not foresee the possibility of reducing it (Kiureghian & Ditlevsen, 2009). For example, the uncertainty associated with betting on the outcome of a single fair coin toss is aleatory (W. E. Walker et al., 2003): Even though one might be perfectly familiar with the process and its possible outcomes, there is a 50% chance of losing. In general, natural and human systems usually involve some degree of aleatory uncertainty, e.g., due to erroneous or erratic behavior (Kiureghian & Ditlevsen, 2009) or due to "the chaotic and unpredictable nature of natural processes" (W. E. Walker et al., 2003).<br>*Synonym*: Inherent randomness (Kabir et al., 2018), variability uncertainty (W. E. Walker et al., 2003) |
| **Deep uncertainty** | Deep uncertainty refers to a high level of uncertainty, with only the extreme of "total ignorance" representing higher uncertainty (W. E. Walker et al., 2013). The presence of deep uncertainty means that one is able to enumerate several plausible alternatives without being able to rank them in terms of likelihood (W. E. Walker et al., 2013). Furthermore, deep uncertainty exists in situations where it is unclear or controversial which models are best suited to describe interactions among a system's variables, how to describe these variables using probability distributions, and/or which system states are desirable (Lempert et al., 2003). |
| **Scenario** | A scenario is a generally intelligible description of a possible situation in the future, based on a complex network of influence-factors (Gausemeier et al., 1998). This description comprises a sequence of events following a well-defined chronological order. Each identified scenario produces a set of consequences, which depends on the |

initiating event, the concerned critical infrastructure, and its geo-organizational context (Setola et al., 2016).

## 3.2 System maintenance

The following terms refer to system properties which are frequently used in the context of infrastructure maintenance.

| | |
|---|---|
| **Safety** | Safety is a state of control over risk of harm or other undesirable outcomes imposed unintentionally (e.g., natural hazardous events, accidents), (Aven et al., 2018; Maurice et al., 1997). |
| **Security** | Security is a state of control over risk of harm imposed intentionally by others (Aven et al., 2018) (e.g., terrorist attacks). |
| **Reliability** | The ability to meet an acceptable or the required performance level over extended periods of time (Beyza et al., 2020), even under unfavorable operating conditions (involving disruptive events of any sort), e.g., "keeping the lights on" in a power system context (NIAC, 2010). In contrast to safety, reliability is always related to specifications regarding the required performance (what is acceptable?) (N. G. Leveson, 2016).<br>*Synonym*: Business continuity (economics context, (BCI, 2013)) |
| **Sustainability** | The term sustainability was first coined in the context of forestry, where it described an approach of not taking more resources out of the system than can be restored, such that a steady resource supply is guaranteed on the long term (see, e.g., Grober (2013) and Spindler (2013) who refer to the book 'Sylvicultura Oeconomica' by Hans Carl von Carlowitz published in (1713) as the first documented use of the term). In the context of social science, sustainable development is characterized as "development that meets the needs of the present without compromising the ability of future generations to meet their own needs" (WECD, 1987), incorporating specifically the concepts of "needs", "intra-generational justice" and "limitations". Examples following the idea of sustainable development are circular economy, material and energy efficiency, end-of-life recovery, and environmental emissions (acatech, 2014; Fiksel, 2003). Sustainability highlights the interconnectedness and interdependence of environmental, social, and economic aspects.<br>The relation between sustainability and resilience is still under discussion (Marchese et al., 2018). Some authors argue that to achieve sustainability, high robustness, adaptation, and learning capacities are needed (acatech, 2014; Fiksel, 2003), and thus resilience is a part of the sustainability concept. However, it can also be argued the other way around, that resilience is the ultimate goal and sustainability contributes to reaching this objective, i.e., sustainability is a part of the resilience concept (S. R. Carpenter et al., 2012; Marchese et al., 2018). Finally, some authors view sustainability and resilience as two complementary concepts, which lack hierarchical order (Marchese et al., 2018). |
| **Resilience** | Resilience is the ability of a system to deal with the impacts of unspecific and possibly unforeseen disruptive events. This ability |
|  | depends on the availability and sophistication of a diverse set of skills and strategies, i.e., the resilience capacities. For example, resilience is enhanced by the capacity to anticipate, resist, absorb, respond to, and recover from negative impacts (Carlson et al., 2012), carry out its original functions (NIAC, 2010), and adapt in response to lessons learned from past experience or changed circumstances (Francis & Bekera, 2014). The core concept of resilience is "the recognition of our ignorance; not the assumption that future events are expected, but that they will be unexpected" (Holling, 1973). The goal is thus to strengthen the system capacities to better deal with future events, whatever form they might take; i.e., without specifying the characteristics of the potential disruptive event in advance. Resilience is less a given, static characteristic of a system, but rather a set of skills and coping strategies that can be actively acquired (Anholt & Boersma, 2018); this means resilience can change over time. Based on the above definition, resilience thus describes a set of various useful skills. In our view, resilience is therefore a synonym of "resilience capacities", while the resulting resilience processes are not part of the resilience term (see Discussion section 4.1 for details). Resilience can be estimated through proxy indicators, as proposed by, e.g., Hollnagel (2011a) or Shirali et al. (2013).<br>*Synonym*: Resilience capacities, resiliency (Braun et al., 2020), genotype of resilience (Kanno et al., 2019) |
| **General resilience** | A type of resilience. General resilience presents the fundamental, underlying resilience capacities of a system (S. R. Carpenter et al., 2012; B. H. Walker & Pearson, 2007). It influences how well a system will be able to react in any circumstance. For example, general resilience describes the overall capacity of a city to persist in a rapidly and unpredictably changing world, without specifying what kind of disruptions occur (B. H. Walker & Pearson, 2007). |
| **Specified resilience** | A type of resilience. Specified resilience describes the capacity of specific characteristics or functions of the system to handle specific types of disruptive events (S. R. Carpenter et al., 2012; B. H. Walker & Pearson, 2007). It is also described as the "resilience of what to what" (S. Carpenter et al., 2001) or as the resilience "regarding what" and "against what" (Tamberg et al., 2022), for example, the resilience of crop production to variation in rainfall, or a power system's total production against extreme wind.<br>Specified resilience thus describes the reaction of the system to a specified class of stressors (or a single stressor), with regard to a number of specified performance indices (or a single performance index), in contrast to general resilience, which describes the reaction of the system as a whole to all imaginable stressors. It has been suggested that due to this restriction, specified resilience can be quantified, while general resilience is difficult to assess (S. Carpenter et al., 2001; B. H. Walker & Pearson, 2007). Accordingly, many instances of specified resilience can be found which often either explicitly specify the "to what" or "against what", e.g., seismic resilience (Bruneau et al., 2003), climate resilience (Kahiluoto et al., 2019), hurricane resilience (Ouyang & Dueñas-Osorio, 2014) or flood resilience (McClymont et al., 2020), or the "of what" or "regarding |
| | what", e.g., water resilience (Simpson et al., 2020) or crop production resilience (Zampieri et al., 2020). However, depending on how strictly the type of disruptive events is specified, this type of resilience conceptually approaches risk management; as resilience should in principle be indifferent towards the form of disruption. |
| **Engineering resilience** | The terms engineering and ecological resilience were coined by Holling in order to distinguish between two different perspectives on resilience commonly applied in the ecological literature (Holling, 1996). According to Holling, engineering resilience refers to the stability of a unique stable state which is composed of its resistance against disruptions and its speed of return to this equilibrium after a disturbance (not to be confused with "resilience engineering", see section 3.7). However, later on, engineering resilience has often been equated with the definition by Pimm (1984) who states that resilience is simply the speed at which a system returns to its long-term equilibrium after a disruption (see, e.g., Herrera (2017)), while resistance is another distinct aspect of a system (see also Pimm et al. (2019)). It is important to note that many (if not most) ecological – especially empirical - works use the term resilience in line with the definition by Pimm, i.e., they equate resilience with what we call rapidity in the context of infrastructure resilience (see, e.g., Hoover et al. (2014), Donohue et al. (2016) or Hillebrand et al. (2018)). Nowadays, the term engineering resilience is sometimes used to refer to frameworks which estimate resilience based on a performance curve (such as in Figure 1), i.e., frameworks which consider resilience as a result- or outcome-oriented concept (see, e.g., Rus et al. (2018) or Asadzadeh et al. (2017)). *Synonyms*: Stability (Holling, 1973), resilience (Pimm, 1984) |
| **Ecological resilience** | As for engineering resilience, the definition of ecological resilience is rooted in dynamical system theory. However, its central premise is the assumption that ecosystems are usually far from any stationary state and, moreover, better described by multi-stable complex systems (Holling, 1973), e.g., systems possessing more than one possible long-term behavior. In this context, ecological resilience has been equated with the capacity of a system to stay in its current basin of attraction, a capacity which is put together by several subcategories (Dakos & Kéfi, 2022; Mitra et al., 2015; B. Walker et al., 2004). Generally speaking, ecological resilience is defined as "the capacity of a system to absorb disturbance and reorganize while undergoing change so as to still retain the same function, structure, identity and feedbacks" (B. Walker et al., 2004). This definition emphasizes the variability and intrinsic self-organizing capability of ecosystems which are necessary for its persistence (Comfort et al., 2010; Folke, 2006). In this regard, and in its understanding of resilience as an emergent system property or ability, the notion of ecological resilience has been very influential and shaped the way in which resilience is understood in various fields (see, e.g., Folke (2006) or Zampieri (2021)). *Synonyms*: Resilience (Holling, 1973), systems resilience (McClymont et al., 2020) |

| | | |
|---|---|---|
| | **Social-ecological resilience** | An extension of the ecological resilience perspective which has been developed in view of social-ecological systems (Folke, 2006; Folke et al., 2016), i.e., systems of inseparably intertwined social and ecological components (Berkes & Jolly, 2002). According to the social-ecological perspective, resilience does not only concern a system's ability to persist (ecological resilience) but puts special emphasis on its adaptability and transformability (e.g., S. Carpenter et al. (2001), Adger (2006) or Folke (2016)). In this sense, resilience is not seen as a static characteristic but as being subject to a dynamic process of continuous development (due to adaptation and transformation) which requires managing resilience for aspects like flexibility, anticipation (forward-looking perspective, e.g., Folke (2016)) and the enablement of social learning (e.g., Cutter (2016b)) – see S. Carpenter et al. (2001) or Elmqvist et al. (2019) for illustrations of the dynamic nature of the concept.<br><br>Today, engineering and social-ecological resilience are often referenced as the two central and opposing perspectives on resilience (e.g., Asadzadeh et al. (2017), Jesse et al. (2019), Liu et al. (2022), Rus et al. (2018) or Butler et al. (2014)) or as the two endpoints of a spectrum of resilience interpretations (Engle, 2011), with the former referring to an understanding of resilience as the outcome following a disturbance (i.e., the return to an equilibrium) and the latter to a dynamic concept with the adaptive capacity as its centerpiece. While the social-ecological resilience perspective is quite popular among contemporary academic resilience definitions, practical applications often still rest upon an understanding reflecting the engineering resilience perspective (Engle, 2011).<br><br>*Synonyms:* Adaptive resilience (Jesse et al., 2019; Liu et al., 2022; Simmie & Martin, 2010), complex adaptive systems resilience (McClymont et al., 2020), socio-ecological resilience (Cretney, 2014) |
| | **Community disaster resilience** | The resilience of a community or any geographically bounded social unit (community resilience: resilience of which system), e.g., a neighborhood, city, county, region or nation, to dangerous hazards and the corresponding disasters (disaster resilience: resilience to what), e.g., earthquakes, hurricanes, tsunamis, pandemics or terrorist attacks (Burton, 2014; Cutter et al., 2008; National Academy of Sciences, 2012; Norris et al., 2008). Usually, the terms community resilience and disaster resilience are used synonymously, although they can refer to something slightly different, e.g., the disaster resilience of a particular infrastructure (H. Cai et al., 2018; Cimellaro et al., 2010).<br><br>In accordance with the variability in the understanding of the resilience concept (e.g., rather complying with the engineering, ecological or social-ecological perspective) and the diverse contexts of applications, a variety of measurement frameworks have been proposed (Asadzadeh et al., 2017; H. Cai et al., 2018; Ostadtaghizadeh et al., 2015; Rus et al., 2018). An important aspect reflected in many frameworks is the understanding of resilience as a multidimensional concept (Bruneau et al., 2003; Burton, 2014; Cutter et al., 2010; Parsons et al., 2016). Accordingly, its evaluation requires the incorporation of inputs from different domains and disciplines, |
| | | e.g., health science, engineering, economics and geography (Cutter, 2016a). A particularly popular approach which allows the integration of multiple dimensions is a so-called composite indicator which combines a number of indicators which are assumed to be essential for building the required system capacities.<br>*Synonyms:* Community resilience (Norris et al., 2008), disaster resilience (Cutter, 2016a) |
| | **Resilience dimensions** | Especially when applied to complex managed systems, such as critical infrastructures, cities or communities, the concept of resilience incorporates multiple interrelated dimensions or domains which contribute to the overall resilience of the system (Bruneau et al., 2003; Cutter, 2016a; Pagano et al., 2018; Qin & Faber, 2019). Although the number of proposed dimensions varies (Cantelmi et al., 2021; Cutter et al., 2010; Guo et al., 2021), the most commonly applied dimensions, and the ones being considered the most essential, are still the four dimensions originally proposed in the seminal work by Bruneau et al. (2003), i.e., the technical, organizational, social and economic dimension (also known as the TOSE dimensions, see below). |
| | **Technical resilience** | The technical dimension of resilience or the resilience of the engineered system. It refers to the resilience capacities or abilities of the physical system or the built environment (Guo et al., 2021).<br>*Synonyms:* Technological resilience (Mottahedi et al., 2021), infrastructure resilience (Cutter et al., 2010; Opdyke et al., 2017) |
| | **Organizational resilience** | The organizational dimension of resilience or the resilience of the owner organization of a system (Mottahedi et al., 2021). This includes all people involved in the management and maintenance of an organization, e.g., managers, personnel, and operators, and depends for example on their problem-solving mentality, training, experience, communication skills and problem-solving mentality as well as on the overall organizational culture and established management processes (Gover & Duxbury, 2018; Pagano et al., 2018).<br>*Synonym:* Governance dimension of resilience (Opdyke et al., 2017) |
| | **Social resilience** | The social dimension of resilience or the resilience of the society in which the system exists (Mottahedi et al., 2021). It refers to the capabilities of social groups or communities to deal with the impacts of disruptive events (Adger, 2000). |
| | **Economic resilience** | The economic dimension of resilience (Guo et al., 2021). It refers to the ability of a system to minimize the financial losses associated with a disruptive event. |
| | **Hard resilience** | Refers to structural, technical, mechanical, and cyber qualities and capacities of a system (Kahan et al., 2009), i.e., the "hard" dimensions of resilience. |
| | **Soft resilience** | Refers to resilience capacities related to community, and society, emphasizing human needs, behaviors, and relationships (Kahan et al., 2009), i.e., the "soft" dimensions of resilience. |
| **Fragility** | | A fragile system significantly decreases in performance when stressed beyond a certain level/threshold (Taleb & Douady, 2012). Fragile systems can be contrasted with resilient systems, whereby consequences of disruptions are much less severe in the latter (Aven, |

|  |  |
|---|---|
|  | 2015a). Fragility thus can be described as the lack of capacities and resources needed to provide basic functionality in the face of changes and disruptions (Manyena & Gordon, 2015; OECD, 2008). Similar to the concept of resilience, the definition of fragility and its' applications are still much debated (Manyena & Gordon, 2015; OECD, 2008). |
| **Brittleness** | Brittleness can be seen as a kind of fragility. A brittle system suddenly collapses when pushed past a threshold by disruptions (Woods, 2015), instead of adapting to the new circumstances, analogous to how materials under stress can experience brittle failure (Woods, 2016). The term is coined and mainly used by Woods (2015, 2016). In the context of infrastructure resilience, it is not widely used. We searched for articles in web of science using the keywords "resilience AND brittle" and found 64 results. All but three were from the field of material science, construction mechanics, or ecology (date of search: 20.04.21, web of science). Two articles mention brittleness in the context of energy or transportation systems (Dias, 2015; Gomes et al., 2009), but they lack a clear definition of brittleness.<br>*Opposite*: Graceful extensibility (Woods, 2016). |
| **Anti-fragility** | The ability to benefit from stress, and potentially increase in performance in reaction to stress (Aven, 2015a; Taleb & Douady, 2012). Anti-fragility has been proposed to be one out of a group of three possible system states, where robustness (defined as stress tolerance without decline in performance) and fragility are the other two optional system states (Taleb & Douady, 2012). As an example of anti-fragility, Verhulsta (2014) presents the aviation industry, which is able to decrease the number of fatalities each year via analysis of and learning from past disruptions. The author contrasts this "antifragility" to "resilience", which in his definition is merely concerned with returning to the same state as before the disruption, without improvement. In contrast to graceful extensibility, anti-fragility thus describes systems which increase performance due to disruptions. |
| **Vulnerability** | Describes how much a system's performance could be reduced by a specific disruptive event, i.e., how strong the negative impacts of the disruptive event would be (Aven et al., 2018; Fischer et al., 2018; Häring et al., 2017). According to Zio (2016), this "weakness" of a system comprises two aspects: (1) the susceptibility to destruction and (2) the incapacity to reestablish stable conditions. However, the definition of the term differs in other fields, e.g., in the security context, vulnerability refers to the likelihood of a negative impact of disruptive events (McGill et al., 2007), in socio-ecological systems it describes the susceptibility to potential harm induced by changes and lack of adaptability (Adger, 2006) and in the field of disaster risk reduction, the term is heavily disputed (Manyena, 2006).<br>While the exact relation between vulnerability and resilience is still discussed, there is consensus that resilience is one of several ways to reduce vulnerability (Jhan et al., 2020; Mottahedi et al., 2021; Rose, 2007); another strategy is mitigation (Rose, 2007) or, more specifically, protection (one aspect of mitigation, see section 3.3). |

## 3.3 Risk management strategies: Minimizing known disruptions before they occur

These strategies aim to reduce the occurrence probability or impact of specific, known disruptions, and are thus part of the traditional risk management processes.

| | |
|---|---|
| **Risk management** | A process that establishes, examines, weighs and conducts plans and actions to identify and monitor hazards/threats, and to avoid them or reduce their impacts, such that the system continues to function (business continuity) (Linkov et al., 2014; National Academy of Sciences, 2012). Focus on high-risk disruptive events, such as foreseeable, single, sudden, high-probability, or high-intensity disruptions.<br>*Includes*: Prevention, mitigation, protection, and absorptive capacity measures (see below and section 3.4; e.g., dams, floodways, early warning systems (National Academy of Sciences, 2012)). |
| **Mitigation** | Reducing the risk due to future disruptions, by targeting the probability of occurrence and/or by minimizing the associated negative consequences of a disruptive event (NIBS, 2019; Rose, 2007; Rose & Dormady, 2018). Mitigation is sometimes seen as part of resilience, e.g., Ouyang and colleagues assign mitigation to the resistant capacity (Ouyang et al., 2012).<br>We propose that based on the determinant of risk (source, exposure, vulnerability) which is targeted, three different classes of mitigation strategies can be distinguished: Prevention, avoidance and protection (see below and Figure 4). While we suppose that this distinction is particular helpful to classify different mitigation measures, it should be noted that often a single measure cannot exclusively be assigned to one of the three classes.<br>*Includes*: Prevention, avoidance, protection<br>*Synonym:* Risk reduction (Aven et al., 2018; Smith & Brooks, 2013) |
| **Prevention** | Reducing risk by targeting the risk source (hazard/threat). In general, prevention can be achieved by manipulating the properties of the risk source (e.g., reducing the probability of severe weather events by climate mitigation), by removing it altogether or by intercepting the pathway from a risk source towards the realization of a disruptive event (Aven et al., 2018). The main scope of prevention is reducing the probability of the occurrence of specific disruptive events (Fischer et al., 2018; Häring et al., 2016; Ouyang et al., 2012). It should, however, be noted that preventive measures can also target the severity of occurring events (which would affect consequence rather than probability).<br>*Synonym:* Hazard intervention (Birkmann et al., 2013)<br>Examples: Routine maintenance, climate mitigation (actions to reduce greenhouse gas emissions to limit global warming, climate adaptation could in contrast refer to avoidance or protection measures (IFRC, 2020)), regulate/prohibit gun possession to prevent gun violence |
| **Avoidance** | Reducing risk by reducing the exposure to the corresponding risk source, e.g., not carrying out the process that may lead to the hazard (Aven et al., 2018; Smith & Brooks, 2013) or by avoiding places or areas where exposure is expected to be high (BBK, 2011). While the main focus of avoidance is minimizing the potential for negative |

| | consequences due to disruptive events, it also affects the probability of incidents becoming (severe) disruptive events in the first place (there is no disruptive event without exposure). *Examples*: Evacuating a building in case of a pending emergency in order to avoid human casualties, not building a facility within an earthquake- or flood-prone area, not moving to a region where drinking water is scarce, removing the risk of armed robbery by paying employees via electronic banking rather than cash (Smith & Brooks, 2013). |
|---|---|
| **Protection** | Reducing risk by reducing the vulnerability of a system, i.e., reducing the impact or consequence in case of a certain exposure to a certain disruptive event (a realized threat) by adjusting system properties (DIN EN ISO 22300:2018; Häring et al., 2016). This could also mean reducing the probability that an event leads to any negative consequences. One way of achieving protection is by preparing contingency and crisis plans which can be utilized in the event of an emergency, crisis or disaster situation (Park et al., 2013; Petersen et al., 2020). Accordingly, protection provides a linkage to reactive response measures which are part of crisis or disaster management (see below).<br><br>In contrast to absorption (resilience-related, see section 3.4), many protection measures are clearly targeted to specific, identified disruptions, for example elevating buildings to reduce the chance that they will be flooded or moving electrical transmission lines underground to better resist wind and ice loads (NIBS, 2019). Nevertheless, since protection directly targets the system or its capacities, it is the mitigation strategy which is most closely connected to resilience - it exhibits some overlap with the absorptive capacity respectively with measures for its enhancement (see below and section 4.1.3).<br><br>*Examples:* Building dams or walls, carrying a shield or an umbrella |
| **Acceptance** | Risks can only be mitigated up to a certain point (Smith & Brooks, 2013). Acceptance describes the informed decision to accept the remaining or residual risk (BBK, 2011; DIN EN ISO 22300:2018). Accordingly, it is primarily concerned with judging identified (i.e., known) risks (Aven et al., 2018): A risk is deemed to be either acceptable or unacceptable, in the latter case, it has to be (further) mitigated. |
| **Crisis management** | All processes which aim at repelling the emergence and reducing the impact of crises and disasters, i.e., crisis management usually involves multiple actors with different responsibilities, e.g., policy makers, risk managers, system operators or technicians (Boin et al., 2018). Although the scope of crisis management is sometimes focused on the crisis response (Williams et al., 2017), most definitions follow a broader understanding of crisis management which incorporates activities before, during and after the emergence of a crisis (Branicki, 2020; Bundy et al., 2017; Christensen et al., 2016). These pre-crisis, during-crisis and post-crisis phases are often further specified to highlight the different tasks which are addressed over the course of the full crisis management cycle, e.g., risk assessment, prevention, preparedness, response, recovery and learning (Pursiainen, 2017). |

Traditionally, crisis management has been deeply rooted in risk management as it heavily relied on analyzing, preventing and preparing response plans/routines and special organizations for known and/or expected situations/scenarios, i.e., this applies to rather recurrent crises, like floods, earthquakes or partial infrastructure failures, which follow familiar patterns to a certain degree (Boin et al., 2018) or simply the most recent severe crisis (Boin & 't Hart, 2010). However, this risk-based approach to crisis management is of limited use for informing actors on how to handle formerly unknown, unprecedented crises (Boin & 't Hart, 2010; Lagadec, 2009). As a consequence, many authors have claimed that a holistic crisis management should incorporate both risk-based and resilience-based management approaches (Boin et al., 2018; Boin & McConnell, 2007; Park et al., 2013).

*Synonym:* Disaster management (we do not distinguish between crisis and disaster management as the broad understanding of crisis management involves both preventing incidents from becoming crises and crises from becoming disasters), catastrophe management

## 3.4 Absorptive capacity: At the beginning of a disruption

This capacity aims to limit the immediate impact of disruptive events on the system. The following table contains sub-capacities or design principles that contribute to the absorptive capacity.

| | |
|---|---|
| **Absorptive Capacity** | The capacity of a system to reduce the initial adverse effects of and to continue to function after a disruptive event (Aven et al., 2018; Jackson & Ferris, 2013). Absorptive capacities aim to ensure persistence of the system (Béné et al., 2012). Absorptive capacity includes aspects that manifest automatically or with little effort (Rose & Dormady, 2018; Setola et al., 2016; Vugrin et al., 2011), in contrast to restorative and adaptive capacity. In contrast to protection, absorption measures are scenario-unspecific and strengthen the general, overall ability of the system to withstand any disruptive event. Perfectly successful absorption is reflected in the performance curve by the absence of a decline (Wied et al., 2020). *Synonym*: Static economic resilience (Rose, 2007), absorbability (H. Xu et al., 2020) *Associated process*: Absorption *Includes*: Robustness, redundancy, diversity, resourcefulness, situation awareness, monitoring and modularity. *Examples*: Safety margin, excess or buffer capacity (Francis & Bekera, 2014; Rose & Dormady, 2018), usage of back-up electricity generators, stockpiling critical materials, as well as importing goods that are in short supply within the affected region, and reserve margins (Rose & Dormady, 2018) |
| **Robustness** | The ability of a system to withstand the immediate impact of a disruptive event. Measures to enhance robustness decrease the impact of disruptive events and are put into place prior to an event (NIAC, 2010; Vugrin et al., 2011). Robustness is one of the most frequently |
|  | used terms to describe resilience capacities (Mottahedi et al., 2021). Robustness can be described by how much of its performance a system can maintain after a disruptive event (Aven et al., 2018; Häring et al., 2017; Mottahedi et al., 2021). Thus, the robustness against a specific disruptive event can be quantified by the minimum performance after the event (Nan et al., 2014), e.g., at least 80% of households have continued power supply after an earthquake (Shinozuka et al., 2004), or, more generally, by the magnitude of a disruptive event divided by the magnitude of the induced initial performance loss (Meyer, 2016). <br> *Synonyms*: Resistance (Carlson et al., 2012), survivability (Mottahedi et al., 2021) <br> *Examples*: Levees that prevent damage by hurricanes to a chemical plant (Vugrin et al., 2011) |
| **Redundancy** | The number of alternate pathways for the system mechanics to operate (Vugrin et al., 2011) or how many independent parts of the system can carry out the same function and thus can replace a part that breaks down to ensure functionality of the system (Kumar et al., 2021). Redundancy increases the absorptive capacity of a system (Mottahedi et al., 2021). <br> *Examples*: A back-up communication system, purchasing input resources from multiple suppliers (Vugrin et al., 2011) |
| **Diversity** | Multiplicity of forms and behaviors (Fiksel, 2003) including variability in the response to disruptions (Elmqvist et al., 2003). Diversity is strongly related to redundancy but aims at avoiding correlated failure of different - possibly redundant - parts of a system (Sterbenz et al., 2010). Often referred to as response diversity in ecological/biological contexts (Elmqvist et al., 2003; Mori et al., 2013) where it is sometimes seen as one aspect of functional redundancy (Kahiluoto et al., 2014; Simpson et al., 2020). Similar to redundancy, diversity is an essential design principle for developing resilient infrastructures (Thier & Pot d'Or, 2020). Diversity comprises for example equipment diversity, human diversity, and software diversity (Thier & Pot d'Or, 2020). <br> *Examples*: Buying components from several manufactures', usage of different protocols for the control of devices (Thier & Pot d'Or, 2020) |
| **Resource-fulness** | The ability of system operators to skillfully manage a disruption as it unfolds. It includes identifying problems, prioritizing what should be done, and communicating decisions to the people who will implement them (Kumar et al., 2021; NIAC, 2010). Implementing these priorities inherently requires the availability of financial and technical resources. In contrast to robustness and redundancy, resourcefulness depends primarily on people, not technology (NIAC, 2010). <br> Sometimes, resourcefulness is seen as a part of the restorative capacity (Mottahedi et al., 2021). In fact, if resourcefulness is high, this can increase both absorptive and restorative capacity (Argyroudis et al., 2020). We rather assign resourcefulness to the absorptive capacity, i.e., targeting an ongoing disruption, acting on a much shorter time scale and implying short-term, temporary changes to the system. It should be noted that what we classify as resourcefulness is referred to as adaptive capacity or adaptability by some authors (see, e.g., Mottahedi et al. (2021), Vugrin et al. (2011) or Francis and Bekera (2014)). In |

| | | |
|---|---|---|
| | | contrast, we define adaptability as the ability for a long-term adaptation, which implies implementing lasting changes in the system to prevent future negative impacts (see section 3.6).<br>*Synonym*: Inherent resilience (Rose, 2007)<br>*Examples*: Good decision-making skills of the system manager, finding new substitutes for critical materials in short supply (Rose & Dormady, 2018) |
| | **Situation awareness** | "Knowing what's going on", building on perception, comprehension, and projection. Perception is the recognition of cues in the environment, comprehension means analyzing the meaning of the multiple pieces of information, and projection describes making predictions about the status in the near future (Endsley, 1995; Mutzenich et al., 2021). Situation awareness increases the short-term ability to deal with acute hazards. Thus, in contrast to anticipation, situation awareness aims to identify the current situation and immediately pending events. It might be argued that situation awareness per definition is concerned with identifying specific disruptions, and thus is not a resilience capacity, but part of risk management. However, the ability to recognize that 'something is going on' and correctly assess and judge this observation's implications on the system is a resilience capacity, as it allows an entity to respond to a formerly unknown and unexpected situation and learn from this process.<br>*Example*: Automated processes in navigation systems to avoid collisions of ships (Engler et al., 2020)<br>*Synonym*: Situational awareness (Stanton et al., 2001) (Note: We prefer the use of the term 'situation awareness' to stress that it is the situation one aims to be aware of, and not something else whose perception could be affected by situational conditions) |
| | **Monitoring** | Monitoring typically describes the continuous/continuing collection of data that informs of the current condition of a system (OECD, 2002). For example, monitoring systems in water distribution networks regularly check water quality (Sankary & Ostfeld, 2017), and railway monitoring systems capture the speed, weight, acceleration and acoustic emissions of trains (Ngamkhanong et al., 2018). Therefore, monitoring can help with the detection of hazards and handling of ongoing disruptions (Jovanović et al., 2018; Ngamkhanong et al., 2018).<br>In contrast to situation awareness, monitoring is not necessarily directed outwards, but emphasizes perceiving the current status and performance of the internal processes of the system. In this sense, monitoring can also refer to the monitoring of a system's own resilience status (see resilience monitoring in section 3.7) which builds the basis for practices like operative resilience management (see section 3.7). |
| | **Modularity** | Modularity refers to system design and describes the extent to which a system is distributed into local rather self-contained compartments or modules (Biggs et al., 2012; S. R. Carpenter et al., 2012). Similarly, in the context of network theory, modularity "refers to the extent to which there are subsets of densely connected nodes that are loosely connected to other subsets of nodes" (Biggs et al., 2012). A modular or |

compartmentalized design allows to maintain residual functionality when parts of a system fail (Jackson & Ferris, 2013) and to contain the spread of disturbances (S. R. Carpenter et al., 2012).
*Synonym:* Localized capacity (Jackson & Ferris, 2013)
*Example:* Operating electrical distribution based on multiple microgrids (Mahzarnia et al., 2020; Schneider et al., 2017), compartmentalization of forested areas due to networks of firebreaks (S. R. Carpenter et al., 2012; Kharrazi et al., 2020)

## 3.5 Restorative capacity: During the disruption

This capacity becomes important once the immediate effects of a disruptive event have manifested and the system is progressing towards reverting these effects. The following table contains sub-capacities or design principles that contribute to the restorative capacity.

| **Restorative capacity** | The capacity of the system to re-establish performance as quickly as possible after a disruptive event. The vast majority of resilience definitions emphasize restorative capacity as a central aspect of resilience (Mottahedi et al., 2021). |
|---|---|
| | It refers to actions that are carried out to revert the effects of the disruption, e.g., sending repair teams out, repairing components by using spare parts, ordering missing spare parts. It is enhanced by contingency plans, competent emergency operations, and the means to get the right people and resources to the right places (NIAC, 2010). Restorative capacity is high if performance is restored quickly and with little resources (Ouyang et al., 2012). In contrast to resourcefulness which often comprises temporary measures to reduce the initial adverse effects of a disruption, restorative capacity typically comprises lasting changes to the system. Moreover, restorative capacity always refers to resolving negative effects of a disruption which manifested in the system, whereas resourcefulness is also concerned with limiting potential secondary implications of a disruptive event and thus enabling fast recovery of performance (Note: A high resourcefulness therefore can improve a system's restorative capacity). The majority of studies define recovery as the return to a normal state of operation, as opposed to few studies which define it as the return to the original state (pre-disruption) of operation (Mottahedi et al., 2021). In general, we prefer the use of the notion 'normal state' or 'desired state' as it comprises the possibility of differing pre- and post-disruption performance levels (see Figure 1). Although it should be noted that what is a normal or desired state (and what not) is system-specific and thus needs to be defined before assessments targeting the restorative capacity are carried out.
*Synonyms*: Rapid recovery (NIAC, 2010), restorability (H. Xu et al., 2020)
*Associated process*: Restoration, recovery, rehabilitation
*Includes*: Rapidity, recoverability
*Example*: High number of spare parts in stock |

| | |
|---|---|
| **Recoverability** | Ability of a system to restore its performance and resilience capacities after a disruption (Mottahedi et al., 2021), for example, in the case of high recoverability, 100% of the original performance are restored. *Synonym:* Repairability (Jackson & Ferris, 2013) |
| **Rapidity** | How fast a system restores its performance after a disruption. This can for example be quantified by the slope of the performance (Mottahedi et al., 2021; Nan et al., 2014), e.g., 10% of performance restored per day; or full restoration within 3 days after earthquake (Shinozuka et al., 2004). *Synonyms*: Recovery slope, engineering resilience (B. Walker et al., 2004), resilience after Pimm (Pimm, 1984; Pimm et al., 2019) *Example*: Fast recovery due to small distance to repair team |
| **Graceful degradability** | The ability of a resilient system to decrease in performance slowly instead of abruptly, when the system capacities needed to cope with a disruption are exceeded (Wied et al., 2020). This enables the system to temporarily operate in a safe but degraded state, and thus can help to return to normal operation after the disruption (Bie et al., 2017). It is also defined as the ability of a system to return to an acceptable state after a disruption, as opposed to a perfectly robust system which would return to the optimum state (Andersson et al., 2020). *Synonym*: Weak robustness (Andersson et al., 2020; Schmeck et al., 2010) *Associated process*: Graceful degradation, graceful deterioration (Wied et al., 2020) *Example:* Switch to safe operation mode during an emergency |

## 3.6 Adaptive capacity: In-between disruptions

We base our proposed distinction of resilience into three main capacities on two main arguments: (1) Each capacity should be associated with one stage within the resilience cycle in which its associated process is dominant, and (2) each capacity should have a unique effect on the system's ability to deal with disruptive events, which can be extracted from performance curves (see Discussion section 4.1.1). However, other plausible frameworks exist which organize the different capacities according to other classification criteria. The variation in interpretations is particularly high for the adaptive capacity. Some authors use the term adaptive capacity to describe temporary actions which prevent system failure (Mottahedi et al., 2021), however, as this is mainly concerned with buffering the impacts of a present disruption, we refer to this as resourcefulness and group it under absorptive capacity (aims at keeping the initial impact low). Others note that adaptive capacities and restorative capacities work in parallel during the recovery period (e.g., visible in Figure 1 in Klimek et al. (2019)) and can therefore only be distinguished based on the underlying mechanism (not based on information gained from analyzing the performance curve or based on the time when the associated process takes place, see Figure 1 in this document) – with the adaptive capacity being defined as the endogenous ability of the system to adjust itself during the recovery phase and the restorative capacity as the exogenous ability to be repaired by external actions during this phase (Nan et al., 2014; Vugrin et al., 2011). In our understanding, both endogenous and exogeneous ability would be included in the restorative capacity since both aim at restoring performance ("it is not straightforward to distinguish their effects on system performance", Nan et al. (2014)). Another interpretation is that the adaptive capacity differs from other capacities in allowing a system to maintain function during particularly severe disruptive events which require adjustments to the current practices (Rose, 2007; Rose & Dormady, 2018), whereby the maintenance of function can refer to both absorbing or restoring performance. Accordingly, we would assign parts of this ability to the absorptive and restorative capacity.

All of these interpretations describe the adaptive capacity as a situational skill, i.e., the ability to skillfully manage an ongoing disruption in order to minimize performance loss (what we refer to as absorptive capacity) or restore performance afterwards (what we refer to as restorative capacity). In contrast, we see the adaptive capacity as a strategic skill, which is focused on learning from past, successfully overcome disruptions (in line with definitions proposed in, e.g., NIAC (2010) or Rehak et al. (2019)). As such, the associated process (adaption) mainly takes place in between disruptions and has no direct effect on the performance curve. Instead, the adaption takes effect in the other two resilience capacities, i.e., if being successful, it strengthens both the adaptive and restorative capacity of a system (see table below for more details).

The following table contains sub-capacities or design principles that contribute to the adaptive capacity. These capacities play out under normal operation of the system and in-between disruptions, not only during disruptions.

| **Adaptive capacity** | The adaptive capacity of a system describes its' ability to change itself to deal with future disruptions (NIAC, 2010; Rehak et al., 2019). It also means implementing changes in the current practices or policies, and to learn from disruptions, e.g., through revising plans, modifying procedures, and introducing new tools and technologies needed to improve before the next crisis (NIAC, 2010; Toroghi & Thomas, 2020). Thus, the adaptive capacity of a system is mainly determined by the social component (human actions), less by technical characteristics (B. Walker et al., 2004). If adaptive capacity is high, the performance of the system might increase compared to the performance before the disruption (Häring et al., 2017), or stay the same even though pressure on the system increases (Béné et al., 2012). However, most importantly, as a follow-up to the disruption, the resilience itself will increase in a highly adaptive system (Rehak et al., 2019). |
|---|---|
| | The adaptive capacity is a uniquely powerful resilience capacity, whose impact shows in the development of the other resilience capacities, i.e., absorptive and restorative capacity. Compared to absorptive and restorative capacity, adaptive capacity is usually not focused on the current disruption but aims to increase the ability of the system to deal with future disruptions (NIAC, 2010; Rehak et al., 2019). For instance, while the restorative capacity is concerned with re-instating a previous level of performance, and thus overcoming the effects of a current disruption (e.g., by repairing a broken component), adaptive capacity aims to increase the ability to cope with future disruptions by making changes to the current practices (e.g., by using a new material to build future components to be more robust to certain impacts). Only in the case of lasting press disturbances whose intensity is too strong to be fully absorbed, the adaptive capacity is engaged in dealing with an ongoing disruption, i.e., it enables the system to adjust to the new conditions (Béné et al., 2012). However, in our |

| | | understanding, this adjustment is mediated via the absorptive capacity: Adaption strengthens the absorptive capacity which then enables the system to absorb the impact of the press disturbance. Because of the central role the adaptive capacity plays in system resilience, some authors even see resilience as a synonym of adaptability (Norris et al., 2008).<br>*Synonyms*: Adaption capacity, adaptability, adaptivity, re-adjust ability (Toroghi & Thomas, 2020)<br>*Associated process*: Adaption, adaptation, process of learning (S. Carpenter et al., 2001)<br>*Includes*: Anticipation, preparedness (Béné et al., 2012) and flexibility<br>*Example*: High awareness that something might go wrong and frequent updates of emergency plans based on recent events (Setola et al., 2016) |
|---|---|---|
| | **Anticipative capacity** | The ability of a system or an organization to forecast future risks (Francis & Bekera, 2014). In a review of more than 25 definitions of resilience, only three explicitly included the ability to anticipate disruptions (Mottahedi et al., 2021).<br>In contrast to situation awareness, anticipative capacity is not concerned with the current situation and near future, but with events that happen in the medium-to-long term. Anticipation goes beyond prediction and includes analysis of the appropriate course of action in a potential future event (Pezzulo et al., 2008).<br>*Associated process:* Anticipation |
| | **Capacity to prepare** | The ability of a system to take appropriate measures in advance to meet potential adverse events (Francis & Bekera, 2014). For instance, in a critical infrastructure, an important constituent of preparedness could be an established crisis management which is flexible enough to be applicable in various different scenarios. Some authors assign preparation or planning to a separate stage of the resilience cycle, instead of including it in adaption (see Figure 1 or Ganin et al. (2016)). We assume that both adaption and preparation take place in-between disruptions and that preparation is only one nuance of adaption.<br>In contrast to avoidance, prevention, and protection, preparation does not aim to change the risk associated with or impact of a specific disruption, but accepts the risk and strengthens the social and intellectual ability to learn from past disruptions and take appropriate actions. For illustration, consider the following simple example: On a construction site, from time to time, bricks accidentally fall and cause head injuries of workers. In this case, the falling brick is the disruptive event, and resulting injury of a worker is the impact of that event. Prevention could mean to check every day whether any bricks have loosened, reducing the probability of a brick falling (i.e., reducing the probability of the event happening by targeting the risk source). Avoidance could mean using machines instead of workers in areas where bricks are particular likely to fall (i.e., reducing the exposure of workers). Protection could mean distributing helmets to the workers, thereby reducing the severity of the injury (i.e., reducing the immediate impact of the disruption). In contrast, preparation would accept that it is not completely avoidable that workers might get injured for various reasons and therefore hire a doctor to treat injuries on site, enabling the workers to quickly return to work (i.e., accepting |
| | | the possibility of disruptions and their immediate effects, but preparing the system to better deal with these impacts). *Synonym*: Planning capacity *Associated process*: Preparation, planning *Examples*: Fast planning processes (Häring et al., 2016), dynamic adaptive policy pathways (Haasnoot et al., 2013), emergency drills |
| | **Flexibility** | The ability of a system to cope with environmental changes (Schmeck et al., 2010) or changes in the acceptable level of performance (Andersson et al., 2020). Flexibility is closely related to graceful extensibility (see below) as it captures the degree to which a system can be changed and extended in comparison to its original design (Andersson et al., 2020). It entails changing the internal processes of the system. While recoverability describes the ability to return to an initial acceptable level of performance, flexibility describes the ability of the system to change its' criteria of what an acceptable performance level is (see Figure 4 in Andersson et al. (2020) for an illustration). *Synonyms*: Sustained adaptability (Andersson et al., 2020; Woods, 2015) *Examples*: Adaption of a company to new limit values set by politics, or step-wise digitalization of an organization |
| | **Graceful extensibility** | The ability of a system to continuously adapt to increasing stress, as opposed to a brittle system which suddenly collapses if a disruptive event leaves the range of events the system was originally designed to handle (Woods, 2015, 2016, 2018). The term was coined by Woods, who describes the term as a blend of graceful degradation and software extensibility (Woods, 2018), with software extensibility describing a principle in software engineering where the need for future software extensions is already considered early on in the design process. Thus, graceful extensibility describes the ability of the system to shift into a new "regime of performance", i.e., transform, invoking new resources, responses, relationships, and priorities. If graceful extensibility is low, the system loses its ability to adapt to new circumstances when stress continuous to increase (Woods, 2018). Woods (2015) describes graceful extensibility as one of multiple conceptual perspectives on resilience. In contrast, we view graceful extensibility as one aspect of the adaptive capacity which specifically targets circumstances where challenges change and grow. The difference to the concept of antifragility is, that an antifragile system would not only continue to function in reaction to stress, but even improve its performance (Aven, 2015b). *Opposite*: Brittleness (Woods, 2018) *Associated process:* Graceful extension |
| | **Transformation ability** | In contrast to adaptability, which describes incremental changes, transformation ability describes the ability to undergo profound changes to the systems primary structure or functions. For example, high transformation ability could help a region move from an agrarian to a resource extraction economy (Béné et al., 2012). In principle, transformability denotes an extreme expression of the capacity to adapt. *Synonym:* Transformability |

| | | *Example:* Move from fossil to renewable energy sources, e.g., in the power supply network or the road infrastructure<br>*Associated process:* Transformation |
|---|---|---|

### 3.7 Management processes targeting resilience

The following terms describe management strategies that are related to the goal of increasing resilience-related capacities of a system.

| | | |
|---|---|---|
| **Resilience management** | | In contrast to traditional risk management, which hardens the system against specific, known threats (Aven, 2019; Francis & Bekera, 2014), resilience management accepts that not all risks are predictable or identifiable in advance, i.e., "surprise is the new normal" (Linkov et al., 2014; UNDRR, 2019) and not all risks can be avoided (Francis & Bekera, 2014). Therefore, emphasis lies less on prevention and solution of individual, foreseeable crises, but more on learning from and adapting to unwanted circumstances in general. Thus, resilience management is less concerned with the prevention and protection of individual risks, but aims at a holistic increase of the capacities to deal with disruptions as they emerge (Anholt & Boersma, 2018). Also, resilience management explicitly includes multiple, slow-onset, low-probability, and low-intensity disruptions.<br>The main subject of resilience management are the resilience capacities or capabilities (Lichte et al., 2022). In second instance, resilience management is a part of performance management, as the aim of resilience is to enable a system to perform well in the face of various disruptions (Aven & Thekdi, 2018). Resilience management typically includes the quantification and evaluation of resilience, as well as the selection, prioritization and implementation of options for improving resilience (Häring et al., 2017). Some authors highlight that resilience management can benefit from approaches inspired by traditional risk management, such as broadly identifying, judging, and classifying potential disruptions, assessing their probabilities and the uncertainty associated with them (Aven, 2017). Similarly, risk management can benefit from resilience management, e.g., measures that are taken to protect the system from unspecified disruptions (e.g., extending the security margin) will always also increase protection for some specific, known disruptions (e.g., against impacts of extreme weather). |
| | **Operational resilience management** | Resilience management based on quantitative measures that capture and implement the concept of resilience (Ganin et al., 2016), i.e., operationalize it or "put it into practice" (Caralli et al., 2016; S. Carpenter et al., 2001; Herrera, 2017). Several frameworks to operationalize resilience management exist, e.g., Schauer et al. (2021), Caralli et al. (2010) or Lichte et al. (2022). Operationalization of resilience needs to be done in advance – before the system is damaged and undesired consequences occur (Kahan et al., 2009). |
| | **Operative resilience management** | Resilience management on a short time scale, i.e., "live" and under operation of the respective system (Jaroš et al., 2016; Thibodeaux & Favilla, 1996). Operative resilience management is based on |

| | continuously updated, quickly changing information. Operative resilience management can thus be seen as the opposite of strategic resilience management (Dvorakova et al., 2017; Klumpp et al., 2010). While operational resilience management emphasizes the applicability of the resilience framework, operative resilience management emphasizes the short time scale on which decisions to improve resilience capacities are made. As such, operative resilience management focuses on managing the current status of resilience capacities in one specific situation. |
|---|---|
| **Strategic resilience management** | Strategic management describes planning on how to reach long-term goals, in contrast to operative management which takes place on shorter time-scales (Dvorakova et al., 2017; Klumpp et al., 2010). For resilience management, this implies focusing on general resilience, i.e., the part of resilience that describes the general ability to deal with disruptions, independent of the current situation.<br>*Synonym*: General resilience management |
| **Resilience engineering** | A kind of resilience management. The term "resilience engineering" was coined by Hollnagel, who describes it as a process that makes a system more resilient by applying four cornerstones of resilience (Hollnagel, 2011a): anticipation, monitoring, responding (i.e., absorptive capacity), and learning (i.e., adaptative capacity). Resilience engineering enhances a system's ability to maintain functionality under varying conditions (including disruptions) (Madni & Jackson, 2009). According to Hollnagel, resilience engineering, in contrast to what he refers to as traditional safety management, not only analyzes "what went wrong" but also focuses on understanding how a system functions under normal conditions ("things that go right") (Hollnagel, 2011b). Thus, following this definition, resilience engineering aims to identify technologies which support resilience, such as self-healing materials or energy-self-sufficient, automated sensor networks (Linkov et al., 2014). It should however be mentioned that the concept of resilience engineering is heavily criticized by some, e.g., that it is based on a wrong representation of what is common practice in safety management (Aven, 2022; N. Leveson, 2020) or that it hardly represents something new when compared to the theory of high reliability organizations (Haavik, 2021). |
| **Resilience by design** | An approach where resilience is built into the system during the design phase, e.g., by explicitly optimizing resilience design principles (see below) like diversity and cohesion (Fiksel, 2003). The term is used mainly by practitioners. For example, the City of Los Angeles conducted a "resilience by design" program to increase seismic resilience, which enforced mandatory retrofits of pre-1980 buildings and stronger standards for newly built telecommunication towers (Jones & Aho, 2019). Other examples come from the field of green engineering, which maximizes intrinsic characteristics to reduce or eliminate the potential negative effects from disruptions (Anastas & Zimmerman, 2003; Fiksel, 2006). While related approaches like risk-based design (Park et al., 2011) or safety-guided design (N. G. Leveson, 2016) aim at mitigating known risks, resilience by design is mainly concerned with building capacities which enable the system to flexibly deal with potentially unexpected events, which includes the |
| | capacity to adapt to changing conditions and to restore performance once failure was inevitable. In contrast to resilience management, resilience by design is not a continuous, active process throughout the life cycle of a system, but it lays the foundation for it.<br>*Synonyms*: Inherent resilience (Fiksel, 2006), resilience-based design (Park et al., 2011) |
| **Design principles** | Design principles describe potential sources of resilience and thus can help build and maintain resilience (S. R. Carpenter et al., 2012; Jackson & Ferris, 2013). Well-known general design principles for resilient systems include for example diversity, redundancy, modularity, subsidiarity, buffer storages, geographical dispersion (Thier & Pot d'Or, 2020), cohesion (Fiksel, 2003), feedbacks, monitoring, leadership, and trust (S. R. Carpenter et al., 2012).<br>*Synonym:* Resilience principles (Jackson & Ferris, 2013) |
| **Resilience-building system properties** | Design principles are an abstract representation of potential sources of resilience (Engler et al., 2020; Jackson & Ferris, 2013). For example, in a particular system, the general principle diversity might refer to supplier diversity, consumer diversity, energy source diversity, software diversity, employee diversity, etc. Accordingly, in order to improve the resilience of an actual system, the abstract principles need to be transferred into practical implications (Engler et al., 2020) which are associated with specific resilience-building system properties.<br>*Synonym:* Resilience resources (Mahzarnia et al., 2020; Richards, 2016) |
| **Resilience monitoring** | Monitoring in general means the continuous and systematic collection of data on specified indicators to track progress towards achievement objectives and use of allocated funds (OECD, 2002). Data collection methods and tools may include document reviews, observations, surveys, focus group interviews, and key informant interviews (Ngamkhanong et al., 2018; World Bank Group, 2017). Resilience monitoring should be used to inform management and stakeholders of the status of resilience capacities (Béné et al., 2015), based on suitable resilience indicators or indices (see section 3.8). |

## 3.8 Resilience analysis

The following terms are frequently used in evaluation and interpretation of resilience and its effects on system performance.

| | |
|---|---|
| **Resilience processes** | The resilience processes describe the resilient behavior of the system, which results from its resilience capacities. The resilience processes thus show how the resilience capacities are realized or "put into action" at one point in time, shaped by uncertainty, the currently present circumstances and the recent past of the system. As such, the resilience processes should not be confounded with resilience itself, where the resilience capacities are the drivers or causes and the processes are the actions that follow from them.<br>The resilience processes can be inferred based on performance-based resilience indices, such as the area under the performance curve (Kanno et al., 2019). |
| | *Synonym*: Phenotype of resilience (Kanno et al., 2019) <br> *Examples*: Absorption, restoration, adaptation |
| **Resilience cycle** | Conceptual sequence of phases in the unfolding of a disruption, in which some capacities and actions are more important than others (see resilience phases in Figure 1). Typically, it includes a preparation/prevention phase, an absorption/response phase, a recovery phase (Fischer et al., 2018; Häring et al., 2017; Rehak et al., 2019). After the recovery phase, an adaption phase might take place, during which the system learns from the disruption and integrates the lessons learned, making it more resilient towards the next disruption (Rehak et al., 2019). For simplicity, the resilience cycle assumes that disruptions do not overlap, and that the main effects of a disruption on the functioning of the system accumulate in a single, concise time span. It is clear, however, that the resilience capacities can only in part be mapped to the phases. For example, learning (Mottahedi et al., 2021) and anticipation are needed throughout the entire cycle, not only in their respective phases. <br> *Synonym:* Crisis management cycle (Pursiainen, 2017) |
| **Performance curve** | Typically refers to the curve of performance over time (Argyroudis et al., 2020; Jovanović et al., 2018; Mishra et al., 2021; Oboudi et al., 2019; Panteli & Mancarella, 2015a; Taleb-Berrouane & Khan, 2019). It is often illustrated in a conceptualized, simplified manner as a typical "bathtub shape", i.e., a straight line, followed by a steep decrease, a pronounced, longer increase and return to the previous level (see Figure 1). To avoid confusion, we advocate to use the term "performance curve" instead of the frequently used term "resilience curve", as the dependent variable is usually performance, not resilience. |
| **Risk curve** | Risk curves summarize system risk in terms of the likelihood of experiencing different levels of performance degradation in disasters as a function of that performance degradation (Shinozuka et al., 2004). |
| **Fragility curve** | Shows the probability of failure or the probability of exceeding certain damage levels in dependence on a hazard characteristic (Kilanitis & Sextos, 2019; Panteli & Mancarella, 2015a; Schultz et al., 2010). To date, it's mostly used in the context of natural hazards (Schultz et al., 2010). <br> *Example*: Damage probability against peak ground acceleration of earthquakes (Kilanitis & Sextos, 2019), failure probability of power system components as a function of wind speed or rain intensity (Panteli & Mancarella, 2015a) |
| **Indicator** | Indicators are quantitative or qualitative measures which are used to describe a characteristic of a system, for example resilience, risk, social vulnerability (Tate, 2012), or sustainability (Fiksel, 2006). Indicators are used in decision making, building consensus, and to explore the processes underlying certain phenomena (Tate, 2012). |
| **Resilience indicator** | Resilience indicators are typically derived based on interviews and estimate certain resilience capacities or constituents which are assumed to be essential within the respective context, such as top management commitment, learning culture, risk awareness, and flexibility (Cutter, 2016a; Hollnagel, 2011b; Shirali et al., 2013; Storesund et al., 2018). |

| | | |
|---|---|---|
| | **Key performance indicator (KPI)** | Key Performance Indicators (KPIs) measure how well a system performs or functions (Engler et al., 2018; Torres et al., 2020). For example, the operational performance of a power grid could be measured using the KPIs "supplied customers" (Braun et al., 2020) (a measure of productivity) or the "maximum frequency deviation" (Groß et al., 2017) (a measure of power quality, see definition of "performance" in section 3.1). |
| | **Key risk indicator (KRI)** | Key Risk Indicators (KRIs) estimate the possible exposure or loss, i.e., the possibility of future negative effects on the system (Torres et al., 2020). When a KRI reaches an unacceptable value, this can be a signal that countermeasures must be taken to reduce negative impacts on the system (Engler et al., 2018). A single indicator can serve both as a KPI and a KRI. For instance, in a power grid, the "maximum frequency deviation" not only measures power quality but, if it crosses a certain threshold, also initiates counteracting measures like load shedding or the disconnection of generators (Groß et al., 2017; Kruse et al., 2021). |
| **Index** | | An index aggregates several indicators, to summarize a complex phenomenon (Hawken & Munck, 2013; Tate, 2012). To this end, indices usually put several measures in relation to each other, often by using ratios. For example, in economics, an index usually relates a price or quantity to a reference standard. The choice of the aggregation model strongly impacts the robustness and reliability of an index (Tate, 2012), as well as the conclusions that are gained applying it (Bakkensen et al., 2017).<br>*Synonym*: Composite indicator (Tate, 2012) |
| | **Resilience index** | Resilience indices combine several measures to estimate the resilience of a certain system. For example, the resilience index by Argyroudis and colleagues integrates information on the robustness and rapidity of assets for multiple hazards, depending on impact, fragility, and occurrence time (Argyroudis et al., 2020). The resilience index by Eldosouky and colleagues relates the resilience of a system to its maximum resilience, by estimating the probability that the system is able to transition from a critical back to a functioning state (Eldosouky et al., 2021). A large variety of resilience indices describe resilience based on properties of the performance curve (see, for example, Bruneau et al. (2003), Francis and Bekera (2014) or Rose (2007)).<br>An alternative to performance- or outcome-based approaches (e.g., approaches using the performance curve) is to estimate resilience based on system characteristics which are assumed to build resilience (see, e.g., Asadzadeh et al. (2017) or Cutter (2016a)), i.e., corresponding approaches do not rely on the occurrence of a disruptive event but focus on the system's potential to deal with potential events. Most common within the diverse family of corresponding approaches (see, e.g., Cantelmi et al. (2021), Guo et al. (2021), Bakkensen et al. (2017) or Asadzadeh et al. (2017) for some extensive reviews) are resilience composite indicator frameworks which integrate measurable information from various sources (e.g., survey data, demographic data, expert opinion) and aggregate/combine them, often following a specific conceptual hierarchy (see, e.g., Storesund et al. (2018), Jovanović et al. (2020) or Petit et al. (2013)), to obtain indices displaying the resilience of a system based on different, e.g., variables |

(Rehak et al., 2019), themes (Jhan et al., 2020), system functions (Fox-Lent et al., 2015) and/or phases of the resilience cycle (Jovanović et al., 2020).

# 4 Discussion and conclusion

The present glossary represents an integrated terminology of resilience-related terms in the context of critical infrastructures. Since the exact meaning of many of the included terms is disputable, the compilation of a clear and consistent set of terms required making some critical choices regarding the conceptual strands we follow. This selection process was inevitably shaped by our own understanding of resilience and by our aim of targeting critical infrastructures. Accordingly, to clarify some of the choices we made, we conclude this document with discussing what is its centerpiece – our understanding of resilience.

## 4.1 Our understanding of resilience (Why capacities?)

Our understanding of resilience combines aspects of the three arguably most fundamental and influential perspectives on resilience, namely, engineering resilience, ecological resilience and social-ecological resilience (see section 3.2). First of all, the presented framework promotes the use of performance curves for assessing resilience (see section 4.1.1), which is rooted in the engineering resilience perspective (see, e.g., Yodo and Wang (2016) or Asadzadeh et al. (2017)). An underlying assumption is that a system has a clearly definable desired state of performance, which it can leave and return to over the course of a disruption. This fits very well as our framework is intended to be applicable to critical infrastructures which, in their very essence (Petersen et al., 2020), provide essential services to the public, i.e., the desired state can be defined as the state of maintained service.

The intended use of performance curves should, however, not be mistaken with the assumption that resilience can be extracted from a single performance curve, i.e., resilience is not the area under the performance curve, neither is it the minimum performance level and/or the slope of the recovery curve (see also Engle (2011) who points to the discrepancy between the most-common definitions and the use of resilience in practice). What can be seen in a performance curve is the outcome of the interplay between a disruptive event and the initiated system response which is shaped by the three resilience processes (absorption, restoration and adaption). A resilience process does, however, only depict one manifestation of an underlying ability or capacity of the system - resilience itself, on the other hand, is this very capacity (or set of capacities) that enables the system to respond in a preferable way. This notion of resilience as the ability or capacity of a system is rooted in the ecological resilience perspective (Holling, 1973; B. Walker et al., 2004).

The incorporation of the adaptive capacity as one of three main constituents of resilience adds the social-ecological perspective to the framework (Adger, 2006; S. Carpenter et al., 2001; Folke, 2016). This addition emphasizes that resilience is not a static trait but subject to permanent development. Accordingly, an effective manipulation of resilience also requires an understanding of how the modification of system attributes might affect the system's potential for further development – in addition to an understanding of how such changes would affect the current system dynamics. Overall, our understanding of resilience as the ability of a system which is comprised of a set of three capacities has important implications for how we imagine resilience can best be assessed (see the upcoming section 4.1.1), how resilience can be built into a system (see the upcoming section 4.1.2) and how resilience – and not risk - can be managed (see the upcoming section 4.1.3).

### 4.1.1 Why three capacities? (the role of performance curves)

The understanding of resilience as a set of a few essential pillar capacities is in line with the work by many other authors (Francis & Bekera, 2014; Rehak et al., 2019). However, the naming and classification deviates from other frameworks which define their pillar capacities based on a different distinction criterion (Béné et al., 2012; Bruneau et al., 2003; Carlson et al., 2012). For instance, Béné et al. (2012) distinguish three capacities – absorptive, adaptive and transformative capacity - based on the level of change to the system which is required to respond to disruptive events of various intensities. Other frameworks, such as the ones presented by Francis and Bekera (2014), Vugrin et al. (2011) or Biringer et al. (2013), use the same naming as we do, but describe a fundamentally different understanding of the adaptive capacity (see section 3.6). Some frameworks largely match our distinction, both in naming and meaning, but include an additional capacity (Linkov et al., 2014; NIAC, 2010), e.g., one which considers a system's ability to prepare or plan (Linkov et al., 2014) or a system's resourcefulness (NIAC, 2010) – both of which we consider to be sub-aspects of one of the three fundamental pillar capacities (see section 3.6 and 3.4).

So why do we choose this particular set of capacities? In principle, our three main capacities are chosen in a way that should ultimately enable us to verify the benefit and efficiency of implied resilience-enhancing measures. We therefore consider two requirements (Bruneau et al., 2003): Firstly, the set of capacities should represent the ultimate goal of all resilience-enhancing measures (requirement I: goal-oriented approach) and, secondly, the choice of pillar capacities should facilitate measuring resilience (requirement II: measurability). In order to meet the first requirement, our capacities need to cover the entire scope of resilience, i.e., all aspects which distinguish a resilient from a less resilient system. We accomplish this by considering two capacities which cover the two essential aspects of coping with an emerging or ongoing disruption (and which are involved in most resilience frameworks) and complement them with a third more strategic capacity which puts special emphasis on the need for considering long-term development, change and flexibility. The resulting set of capacities consists of (1) the capacity to keep the initial impact of an unspecific disruptive event as small as possible (absorptive capacity), (2) the capacity to recover fast and as completely as possible from disruptions (restorative capacity) and (3) the capacity to learn from disruptions and implement corresponding changes to the system and thus reduce the impact of future disruptive events (adaptive capacity).

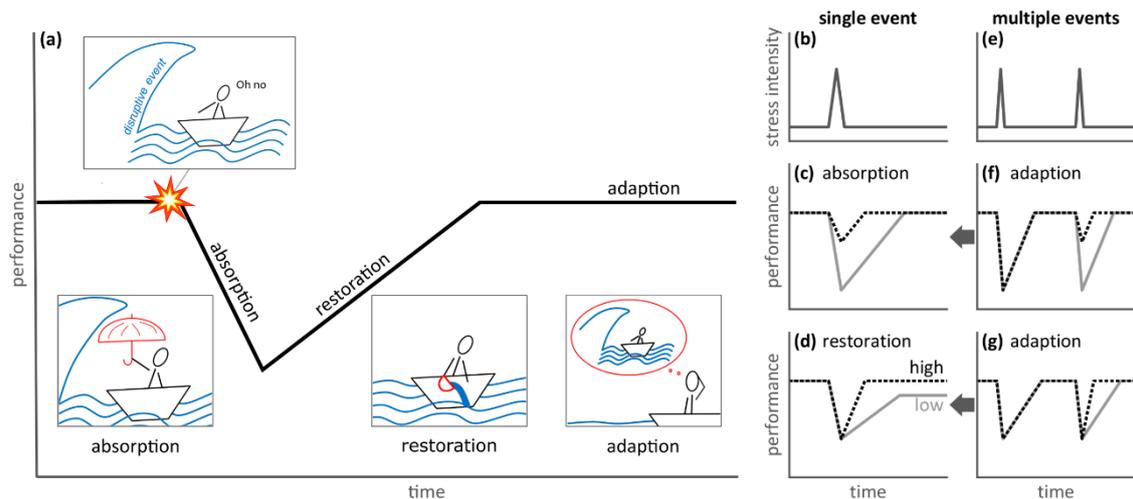

**Figure 2: Effect of the three essential resilience processes.** (a) Visual representation of each process based on one exemplary disruptive event. For illustrative purposes it is helpful that each process can be associated with one stage within the resilience cycle – although in reality the different stages are likely to overlap in time. (b-g) Effect of the three processes on system performance. The effect is depicted in the case of a single disruptive event for absorption and restoration (b-d), and multiple disruptive events for adaption (e-g), in each case for a system with a higher (black dashed line) and a lower capacity (gray solid line).

Furthermore, the three main capacities fulfil our second requirement, i.e., they build the basis for measuring resilience. In this regard, our main consideration was that, in order to assess all three capacities, each of them should be associated with a unique and distinguishable process which is involved in dealing with a disruptive event (see Figure 2a). The improvement of each process should further have a clear imprint on performance curves (see right side of Figure 2), e.g., if one of two systems is able to better absorb the same disruptive event, this should show in characteristic differences between the two corresponding performance curves. For absorption and restoration, some of the sub-aspects we assigned to the corresponding capacity (robustness in section 3.4, recoverability and rapidity in section 3.5) describe how an enhancement would affect the performance curve in the case of a single disruptive event: A better absorption leads to a smaller initial impact (robustness) and thus to higher minimal performance level (Figure 2c); a better restoration increases the magnitude (recoverability) and rate (rapidity) of performance increase after a disruptive event (Figure 2d). For adaptation, the situation is less clear. In the literature, the net gain in performance over one resilience cycle (increase in performance relative to the performance level before the disruption) is sometimes considered a measure of adaption (see, e.g., Häring et al. (2017), Klimek et al. (2019) or Jovanović et al. (2020)). We refuse this notion as we consider the optimization of system performance under normal or stress-free conditions as being out of the scope of resilience (a net performance gain could nevertheless be an indication for the occurrence of adaption). In our understanding, adaption means learning to better deal with future disruptive events. We propose that this learning process can best be depicted as a change in the other two capacities, i.e., the adaptive capacity is a system's ability to adapt its two coping capacities (absorptive and restorative capacity). Accordingly, the impact of an enhanced adaption is not directly visible in a single resilience cycle but only shows in the longer-term development of the other two capacities and their associated processes, i.e., in an enhanced absorption or restoration in a future disruption (Figure 2fg).

### 4.1.2 How to build resilience? (the role of principles)

The use of the three proposed pillar capacities facilitates the assessment of resilience based on performance curves. However, we emphasize here that the regulating screws, which need to be manipulated in order to boost resilience, are the properties of the managed system (the resilience-building system properties). Design or resilience principles (see section 3.7) can provide guidance for choosing appropriate resilience-building measures. We included the principles we believe to be of particular importance in the glossary. Some of them rather refer to "ends" (Bruneau et al., 2003) which describe what a resilient design should accomplish (e.g., robustness, rapidity, anticipative capacity) and some of them rather to "means" which describe how to accomplish these goals (e.g., redundancy, diversity, flexibility). However, all of them help to depict what a resilient system should look like.

In the literature, design or resilience-building principles are often categorized into few distinct groups, in accordance with the capacity or the higher-level goal they promote (Jackson & Ferris, 2013; Rehak et al., 2019). We followed this approach and assigned each principle as a subcategory to one of three pillar capacities (see section 3.4-3.6). However, this assignment is not entirely unambiguous. In fact, that a specific design principle concerns only one of the three capacities is rather the exception than the norm. For instance, a modular design might at first hinder the propagation of an initial harmful impact within a system and thus enhance absorption, but subsequently it can also promote building up the prior performance level – e.g., operating energy distribution grids within multiple smaller balancing areas (microgrids) allows an islanded operation during emergencies which ultimately can also shorten restoration times (Panteli & Mancarella, 2015a). We assume that the same holds true for most resilience-building principles and thus that most principles have several cause-effect-pathways (see Figure 3 for our suggestion); although the actual relations will ultimately depend on the specifics of the system of interest.

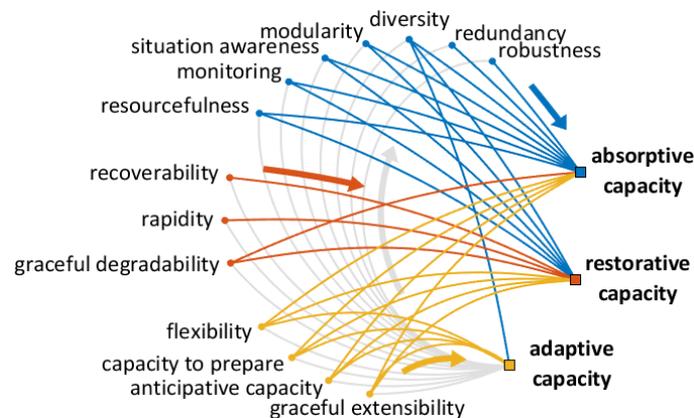

Figure 3: Connection between resilience principles and the three resilience capacities. The colored edges denote which principle is assumed to affect which capacity. Gray edges in the background indicate that the adaptive capacity can take effect by altering any resilience-building system property.

In the actual implementation of resilience principles inter-dependencies can arise (see also Jackson and Ferris (2013)). In the example mentioned above, for instance, a modular design is a necessary condition for an islanded operation of the power grid which allows system operators to switch operation modes - from a global to a local balance between power supply and demand - in case of an emergency. The availability of multiple operation modes ultimately raises the number of options to skillfully handle a disruption. Accordingly, the improvement of modularity allows improving diversity and resourcefulness. Such dependencies are likely to occur since more technical principles like redundancy or monitoring can often build the foundation for more organizational principles like resourcefulness and situation awareness.

Furthermore, the interrelation between different principles can also involve trade-offs. For instance, building underground instead of overhead power transmission lines might improve the robustness of a power grid but, if damaged, also hampers its repairability (Panteli & Mancarella, 2015a). In the end, we cannot expect simple cause-effect relationships between system design and system resilience – especially in complex dynamical systems such as critical infrastructures. Therefore, building resilience always requires a holistic view on the entire system, otherwise one could overlook important trade-offs, dependencies or synergies between different design objectives.

### 4.1.3   Where does it end? (the difference between building resilience and reducing risk)

To highlight the boundaries of the resilience concept and the scope of corresponding management efforts, it is helpful to consider the difference between approaches which aim for reducing risk and those which aim for building resilience. While answers to the question after the difference and relation between the two approaches heavily depend on the consulted author (see section 2.4), an important and widely accepted distinction criterion is the uncertainty or "knowability" associated with the events the two approaches aim to prepare for: While risk approaches help preparing for events which fall into a spectrum of known or familiar patterns, resilience approaches aim at enabling a system to effectively respond to any event, including those which are unexpected and unique (Boin et al., 2018; Park et al., 2013). In fact, it is the growing awareness of the latter class of events which caused the rising popularity of "resilience thinking" in critical infrastructure policies (OECD, 2019; Petersen et al., 2020; Pursiainen & Kytömaa, 2022).

The difference in the targeted events has implications for the nature of strategies which can be followed to build resilience. We have proposed that mitigation or risk reducing strategies can be categorized based on the risk determinant they aim at (see section 3.3): Prevention aims at reducing the probability of severe hazards or threats, avoidance targets the exposure to these risk sources and protection intends to reduce the vulnerability of the system (see left side of Figure 4). Two of these three strategies, prevention and avoidance, are out of the scope of valid resilience-enhancing measures (see right side of Figure 4) as they apply only to formerly known hazards or threats (see also Boin et al. (2018)), i.e., prevention and avoidance are clearly and exclusively risk-based.

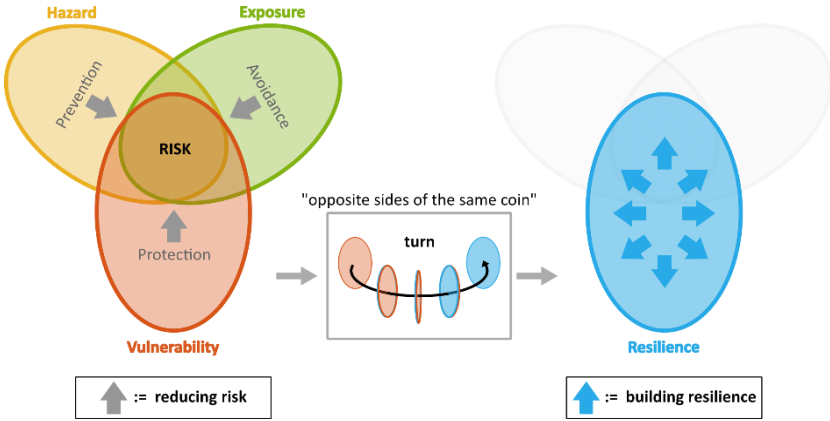

**Figure 4: Difference and overlap between reducing risk and building resilience.** Each risk reducing strategies – prevention, avoidance and protection (depicted by gray arrows) - aims at one of the three risk determinants at the point of expected risk, i.e., where hazard, exposure and vulnerability are anticipated to come together. Building resilience shows some overlap with protection since the two measures act on "opposite sides of the same coin" (IFRC, 2020), i.e., both concern capacities of the system. However, building resilience is independent from any expected risk and thus corresponding measures target the entirety of resilience capacities (depicted by blue arrows pointing at no particular direction). Illustration on the left side is based on Lavell et al. (2012).

In contrast, the third risk-reducing strategy – protection - shows some overlap with approaches for enhancing resilience: Both aim to decrease the impact of disruptive events by implementing changes which strengthen system capacities, protection from the risk perspective (targeting the vulnerability against specific events), resilience enhancement from the resilience perspective (targeting the capacity to cope with any event) – we therefore depicted the two as acting on "opposite sides of the same coin" (IFRC, 2020). Resilience-enhancing measures that are taken to protect the system from unspecified disruptions (e.g., extending the security margin) will always also increase protection against some specific, known disruptions (e.g., against impacts of extreme weather) – i.e., improving resilience can be considered a valid risk-reducing strategy (as done, e.g., in Birkmann et al. (2013) or O'Brien et al. (2006)). In the same manner, it is reasonable to assume that protection measures which are taken with regard to specific threats, such as setting up disaster response plans or undertaking training for specific disaster scenarios, generally strengthens a system's ability to deal with unanticipated disruptions. Ultimately, the effectiveness of the two proactive measures will cumulate in the system's response to an emerging disruptive event. Nevertheless, the conceptual difference between the two opposing perspectives is reflected in a distinct methodological approach. As part of risk management, protection presents one mitigation option for meeting risks which are considered unacceptable (the other two options being prevention and avoidance). Accordingly, it will be incorporated in the process of weighing and prioritizing different risks, deciding which of the identified risks poses the greatest threat, what are their likelihoods of occurrence and what are the most efficient ways to mitigate the most pending ones. Resilience management, on the other hand, is not concerned with treating specific risks, but with maintaining and enhancing the resilience capacities, i.e., the ability of the system to absorb, recover, and adapt to any type of stress under any circumstance. Enhancing resilience is thus the more holistic and system-centered approach to building system capacities, which includes a system's long-term development and its ability to improve itself (explicitly included due to the adaptive capacity). We therefore depicted resilience enhancement as pertaining the complete coin representing resilience, while protection only concerns the part of vulnerability which overlaps with the anticipated threat and exposure (see Figure 4).

In the end, reducing risk and building resilience are conceptually distinct tasks which differ in scope and focus, but both share the common goal of keeping the negative impacts of disruptive events as mild as possible. Therefore we strongly believe that the best way to go when aiming for safe and secure infrastructures is an $R^2$ approach which involves both efforts to reduce risk and efforts to build resilience ($R^2$ equals reducing risk times building resilience): Prevent and/or avoid known threats where possible (with reasonable effort), protect the system against the remaining risk and build resilience in order to be prepared for unexpected or unpredicted events (prediction will inevitably fail sometimes). Such a holistic approach will be capable of dealing with both, the known and unknown (Park et al., 2013) and targeting both the reduction of the probability of occurrence of events and the capability to deal with these events should they arise (Jovanović et al., 2017).

## 4.2 Conclusion (What is it good for?)

It has been argued that science "largely depends upon clear and commonly agreed upon definitions of concepts, and well-defined parameters" (Blokland & Reniers, 2020). While we strongly support the use of clear definitions and the communication thereof, we are also aware that the field of resilience research is way too diverse and divided to be summarized in a unified glossary that pleases everyone. Instead, the goal of this glossary is to present an integrated and consistent terminology which displays one possible perspective on resilience - resilience from a viewpoint we believe is particularly helpful when striving for an operational and operative treatment of the resilience of critical infrastructures. In this regard, it can serve as an introduction for scholars which are new to the field and as a source for scholars sharing a similar scope in their work. By focusing on one specific perspective, this glossary can create a firm conceptual basis for transferring resilience from theory to practice, e.g., when figuring out how to measure, build or manage resilience.

While we focus on presenting one perspective on resilience, we also paid tribute to the diversity and ambiguity surrounding the concept by drawing from a diverse pool of ideas developed in different disciplines, integrating different views where possible and referring to adjoining or contesting views where necessary. In this way, we created entry points for scholars with a different understanding and/or background. Such entry points allow scholars to align the presented definitions with their own understanding and, if they disagree with part of the glossary, to easily adapt it to their needs. The background is that we intend to avoid depriving the resilience concept from being malleable, since its malleability makes much of the concept's appeal within and across disciplinary boundaries (Brand & Jax, 2007; Meerow et al., 2016), which ultimately provides opportunities for transdisciplinary exchange and cooperation (Moser et al., 2019). In the end, by focusing on one perspective while appreciating the diversity surrounding the concept, this glossary provides orientation in the sometimes-confusing field of resilience terminology, it builds the conceptual basis for resilience applications, and it invites a wide group of people into using, discussing, adapting and further developing the presented ideas.

# 5 Acknowledgements

We thank the members of the department for resilience and risk methodology for their valuable input and the insightful discussions; Walaa Bashary, Jens Kahlen, Alexander Khanin, Ingo Schönwandt, and Dustin Witte. Furthermore, we thank Tobias Koch and Kostyantin Konstantynovski for their useful comments regarding an earlier version of this manuscript.